\documentclass[]{aa} 
\usepackage[]{graphicx}
\usepackage{natbib, epsfig, amsmath, amssymb, mathrsfs, txfonts,bbm,xspace,alltt}
\usepackage[utf8]{inputenc}
\usepackage[T1]{fontenc}
\usepackage{color}
\DeclareGraphicsExtensions{.eps, .jpg, .ps}
\usepackage{epstopdf}
\bibpunct{(}{)}{;}{a}{}{,}

\begin{document}

\title{Fine cophasing of segmented aperture telescopes with ZELDA,\\a Zernike wavefront sensor in the diffraction-limited regime} 
\titlerunning{Fine cophasing with ZELDA}

\author{P. Janin-Potiron \inst{1}, M. N'Diaye \inst{1}, P. Martinez \inst{1}, A. Vigan \inst{2}, K. Dohlen \inst{2}, and M. Carbillet \inst{1}}
\institute{Universit\'e C\^ote d'Azur, Observatoire de la C\^ote d'Azur, CNRS, Laboratoire Lagrange, UMR 7293, CS 34229, 06304 Nice Cedex 4, France
\and 
Aix Marseille Universit\'e, CNRS, LAM (Laboratoire d'Astrophysique de Marseille) UMR 7326, 13388 Marseille, France} 
\offprints{Pierre.Janin-Potiron@oca.eu}

\abstract
{Segmented aperture telescopes require an alignment procedure with successive steps from coarse alignment to monitoring process in order to provide very high optical quality images for stringent science operations such as exoplanet imaging. The final step, referred to as fine phasing, calls for a high sensitivity wavefront sensing and control system in a diffraction-limited regime to achieve segment alignment with nanometric accuracy. In this context, Zernike wavefront sensors represent promising options for such a calibration. A concept called the Zernike unit for segment phasing (ZEUS) was previously developed for ground-based applications to operate under seeing-limited images. Such a concept is, however, not suitable for fine cophasing with diffraction-limited images.}
{We revisit ZELDA, a Zernike sensor that was developed for the measurement of residual aberrations in exoplanet direct imagers, to measure segment piston, tip, and tilt in the diffraction-limited regime.}
{We introduce a novel analysis scheme of the sensor signal that relies on piston, tip, and tilt estimators for each segment, and provide probabilistic insights to predict the success of a closed-loop correction as a function of the initial wavefront error.}
{The sensor unambiguously and simultaneously retrieves segment piston and tip-tilt misalignment. Our scheme allows for correction of these errors in closed-loop operation down to nearly zero residuals in a few iterations. This sensor also shows low sensitivity to misalignment of its parts and high ability for operation with a relatively bright natural guide star.}
{Our cophasing sensor relies on existing mask technologies that make the concept already available for segmented apertures in future space missions.}

\authorrunning{Janin-Potiron et al.}
\maketitle

\section{Introduction}
The James Webb Space Telescope (JWST), NASA's forthcoming orbiting observatory, will follow the Hubble Space Telescope and provide new insights into the birth and evolution of galaxies, stars, and planets \citep{CLAMPIN14}. This observatory includes a 6.5 meter diameter telescope with a segmented primary mirror made of 18 hexagonal segments. 
To perform as a single monolithic mirror telescope and provide diffraction-limited images, a wavefront sensing and control system is required to sense and correct for any errors in the optics. {The} segment alignment process \citep[e.g., ][]{ACTON2012,KNIGHT2012,LIGHTSEY2012,GLASSMAN2016} is a long operation including multiple steps needed to pass through commissioning (coarse alignment, coarse phasing, and fine guiding) to maintenance procedure (fine phasing and wavefront monitoring). Segments exhibit misalignment with excursions that are initially much larger than the {observing wavelength and which must be aligned to a} few nanometer residuals.
All of these alignment processes that assume different sensing requirements are mandatory to align the telescope into a high-performance observer. The last procedure is {referred to as the fine phasing process, which} produces a sharp and coherent point spread function (PSF) near the diffraction limit.  

One of the {potential successors to JWST}, the large UV/optical IR surveyor (LUVOIR) is currently studied by NASA and should provide a larger segmented aperture from eight to 16 meters with a primary mirror made up of at least 36 segments \citep{FRANCE16}.
One of the primary science goals of LUVOIR is to directly image and characterize Earth-like planets around nearby stars \citep{CROOKE2016}. Such observations require contrast levels up to $10^{10}$ and wavefront stability down to a few picometer levels, calling for, amongst other features, a precise control of the cophasing errors \citep[e.g., ][]{YAITSKOVA03,LYON2012,STAHL2013,REDDING2014}.
In anticipation of the increase in the system complexity expected with these future telescopes, it is worth exploring new concepts for segment cophasing, and in particular within the diffraction-limited domain. 

Most of the current cophasing sensors are based on existing wavefront sensors that are usually employed in adaptive optics (AO), but thoroughly re-adapted considering that traditional AO sensors assume continuity of the wavefront. 
The recovery of the segment misalignment can be directly obtained from the information in the image plane \citep{LOFDAHL98,DELAV10, MARTINACHE13, POPE2014, JANIN2016}, in a pupil plane \citep[e.g., ][]{CHANAN89, MONTOYATHESIS, ESPOSITO05, DOHLEN06, MAZZOLENI08,  PINNA08}, or in intermediate planes  \citep[e.g., ][]{CHANAN99, CUEVAS00, CHUECA08}. 
Image plane techniques are advantageous especially for space applications since they require a limited amount of hardware. 
However, pupil plane methods bring together the following advantages: (1) the relationship between the sensor and the telescope pupil locations is direct and straightforward; (2) in addition to {segment} alignment (piston and tip/tilt), pupil plane methods are more favorable to access segment figuring errors (defocus, astigmatism, trefoil) and segment higher order wavefront error. 

Among these concepts, we here consider the Zernike wavefront sensor (ZWFS), {which is} based on the phase-contrast method \citep{ZERNIKE34}. These concepts aim to modulate phase aberrations on an unresolved star image with a phase-shifting mask into intensity variations in a pupil plane. Over the past few years, different ZWFS kinds have been proposed in astronomy to address various applications, such as wavefront sensing in AO systems \citep{BLOEMHOF03,BLOEMHOF04a,DOHLEN04}, calibration of the non-common path aberrations in ground-based facilities \citep{WALLACE2011,NDIAYE13,NDIAYE16}, and measurement of the pointing errors or focus drifts in exoplanet imagers \citep{ZHAO14,SHI15}. Such sensors have also been envisioned for the cophasing procedures of segmented aperture telescopes which include coarse and fine alignment regimes \citep{DOHLEN04,DOHLEN06,SURDEJ10,VIGAN11}.

For many years, ZWFS studies have been focused on the coarse phasing regime to address the discontinuous wavefront with segmented aperture telescopes on the ground under atmospheric turbulence. In this context, cophasing errors of a few tens of nanometer need to be detected in the presence of a few microns of wavefront errors that evolve at a rate of a few hundreds of Hz. During the preliminary European-extremely large telescope (E-ELT) studies, the Zernike unit for segment phasing (ZEUS) was proposed in the framework of the European southern observatory (ESO) active phasing experiment \citep[APE, ][]{GONTE2008} to align telescope segments under seeing conditions \citep{DOHLEN06}. On-sky demonstrations showed the ability of this concept to reconstruct a discontinuous wavefront with an accuracy better than 15\,nm rms \citep{SURDEJ10}.

To achieve such a performance, the ZEUS design is optimized with a mask of seeing disk size. This allows the sensor to filter out the low spatial frequency content of the wavefront errors that is dominated by the atmosphere and hence, to enhance the high spatial frequency information that is related to the segment misalignment. The relative piston, tip, and tilt between the segments are then retrieved thanks to a careful analysis of the signal at the border between adjacent segments. While these alignment measurements are limited to half a wave by the ZEUS capture range in monochromatic light, \citet{VIGAN11} have shown that multiwavelength-based strategies enable the iterative calibration with this sensor of piston and tip tilt errors that are larger than one wave.

Despite these encouraging results in the coarse phasing regime, the use of a ZWFS has been little unexplored in the field of fine cophasing. Neat and actively controlled alignment is nevertheless mandatory when instruments yield diffraction-limited images to reach a high, stable wavefront quality in the high-Strehl regime. In particular, this aspect is crucial for high-contrast observations of circumstellar environments with future large observatories. The current implementation for segment phasing ZEUS is well-suited for coarse alignment under seeing conditions. Still, this Zernike sensor cannot address small wavefront errors for two reasons: its seeing-disk-size mask is not suitable to produce an interpretable signal in the diffraction-limited regime and the segment edge-to-edge analysis proves limited in the presence of a faint signal. Sensing a few nanometer cophasing errors in the high-Strehl regime with a ZWFS requires a mask with a size adjusted to the resolution element diameter of the star image but also a revisiting of the sensor signal analysis.

Recently a ZWFS called ZELDA has been proposed to calibrate the residual aberrations in exoplanet direct imaging instruments. This concept was validated with success on a real exoplanet direct imager using the Spectro-Polarimetric High-contrast Exoplanet REsearch (SPHERE) on the very large telescope \citep[VLT, ][]{BEUZIT08}, providing measurements of small aberrations with nanometric accuracy \citep{NDIAYE13, NDIAYE16}. ZELDA uses a diffraction-limited size mask and an algorithm based on the analysis of the whole sensor signal to retrieve residual wavefront errors. Such features make this concept attractive 
for the calibration of segment phasing errors. 

In this paper, we investigate the use of ZELDA for the measurement of piston, tip, and tilt errors in segmented apertures within the diffraction-limited domain. In Section \ref{sec:principle}, we recall the principle of the concept and propose an algorithm to extract the segment phasing errors from the sensor signal. In Section \ref{sec:numerical}, we provide a calibration scheme for segment cophasing in closed-loop operations using our sensor. In Section \ref{sec:results}, we finally discuss the overall performance of the sensor. 

\section{Analytical approach and system response}
\label{sec:principle}

\subsection{Zernike phase filtering sensor}

The ZELDA sensor uses a phase-shifting mask located in the focal plane downstream of the telescope pupil to sense aberrations on an unresolved star image (see layout in \citet{NDIAYE13, NDIAYE16}). 
With a relatively good centering of the focal plane mask (FPM) on the star image, the starlight contributions going through and outside the mask interfere in the relayed pupil plane, yielding a light distribution that is directly related to the phase wavefront errors in the entrance pupil $\phi$, according to the mask characteristics, that is, the diameter $d$ and the introduced phase delay $\theta$.
In the following, $\lambda$ and $D$ denote the wavelength of observation and the telescope aperture diameter. 

As previously stated in \citet[Eq.~(8) of that paper]{NDIAYE13}, the ZELDA relayed pupil plane intensity $I$ can be expressed as
\begin{equation}
\begin{aligned}
    I(\phi)=P^{\,2}&+2b^{\,2}\left(1-\cos\theta\right)
    \\
    &+2Pb\left[\sin\phi\sin\theta - \cos\phi\left(1-\cos\theta\right)\right],
    \end{aligned}
\end{equation}
where $P$ is the amplitude pupil function and $b$ the electric field amplitude diffracted by the FPM in the re-imaged pupil plane.

Assuming a mask angular diameter $d$=1.06\,$\lambda/D$ with a circular pupil gives a value $b$ close to 0.5 over the pupil. In the classical case where $\theta=\pi/2$, the previous equation becomes
\begin{equation}
\label{eq:signal}
    \begin{aligned}
I(\phi)=P^{\,2}+2b^{\,2}+2\sqrt{2}Pb\sin\left(\phi-\phi_0\right),
    \end{aligned}
\end{equation}
where $\phi_0=\pi/4$ represents the reference phase in the absence of aberrations, an inherent property of ZELDA. In the small aberration regime, a first-order Taylor expansion of $\phi$ allows one to retrieve the phase term from the measured intensity \citep{NDIAYE13}. For the estimation of cophasing errors, we introduce no approximation in this equation and we directly work with the re-imaged pupil intensity, keeping the exact expression of $I$ as a function of $\phi$, as shown in Eq. (\ref{eq:signal}).

\subsection{Cophasing estimators}
\label{subsec:analytical}

\begin{figure*}[!ht]
\centering
\includegraphics[trim={2cm 10cm 2cm 0},clip=true,width=0.9\textwidth]{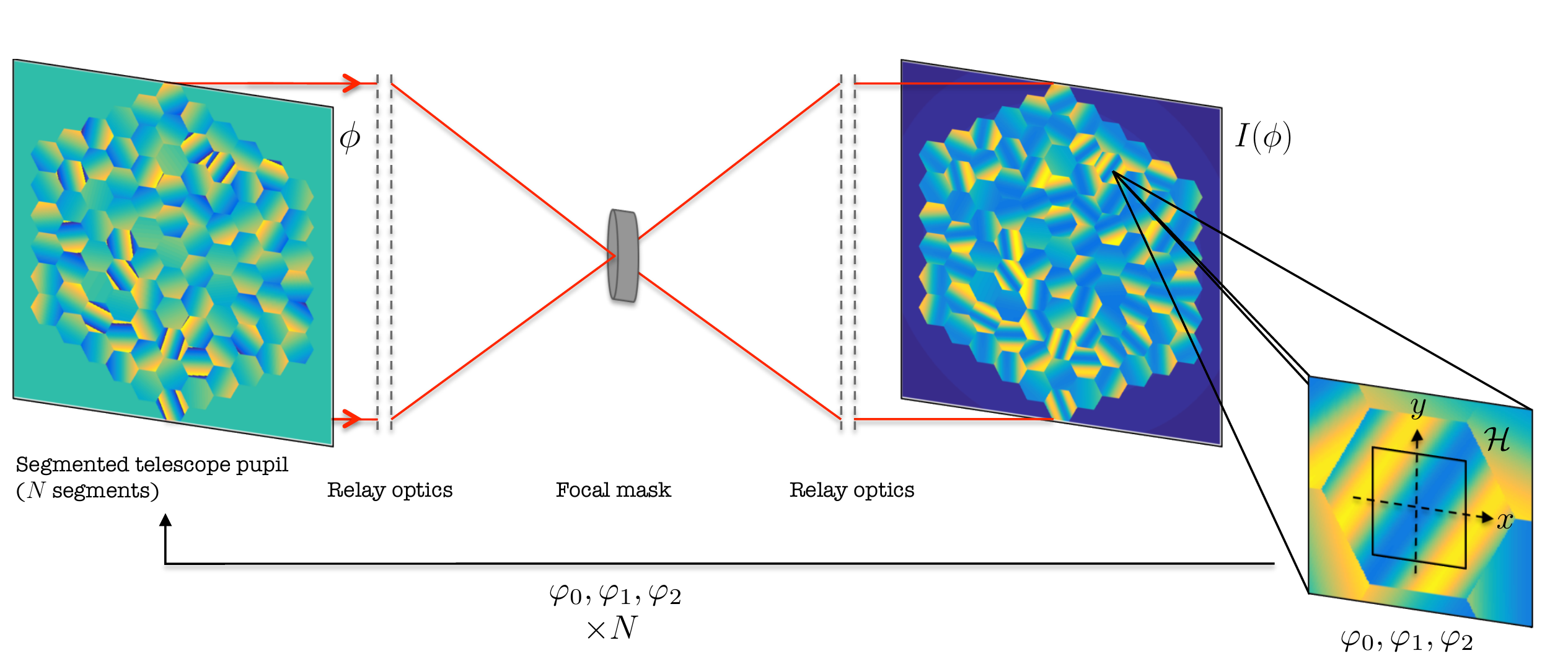}
\caption{Illustration of the ZELDA cophasing sensor principle. From the intensity in the relayed pupil plane $I$, three estimators are evaluated on each segment and the system evolves in closed loop.}
\label{fig:zelda_scheme}
\end{figure*} 

We now define a set of estimators to retrieve the cophasing errors in a segmented pupil made of $N$ segments (see Fig.~\ref{fig:zelda_scheme} in which the axes are defined). Each segment is subject to aberrations of piston (translation along the optical axis), tip, and tilt (rotation around the $\mathbf{x}$-axis and $\mathbf{y}$-axis). Vectors describing these aberrations for all the segments are respectively defined as
\begin{equation}
\begin{array}{cclccccrr}
   \mathbf{p}&=&\left[\right.&p_1 & \cdots & p_n & \cdots & p_{N}&\left.\right], \\
   \mathbf{t}&=&\left[\right.&t_1 & \cdots & t_n & \cdots & t_{N}&\left.\right], \\
   \mathbf{T}&=&\left[\right.&T_1 & \cdots & T_n & \cdots & T_{N}&\left.\right]. \\
\end{array}
\end{equation}
{The segmented telescope is assumed to be a reflective system and the aberrations are expressed as the mirror displacement at the wavefront.} The elements $t_n$ and $T_n$ correspond to the gradients according to the $\mathbf{x}$- and $\mathbf{y}$-axes. The {wavefront error $h_n$ from} the $n^\text{th}$ segment at the position $(x,y)$ is thus given by
\begin{equation}
\label{eq:elevation}
    h_n\left(x,y\right)=p_n+t_n.x + T_n.y \quad \forall (x,y) \in \mathcal{S},
\end{equation}
where $\mathcal{S}$ is the surface defined by a single segment. On the $n^\text{th}$ segment, $\phi$ is related to $h_n$ by
\begin{equation}
\label{eq:elevphase}
    \phi = \frac{2\pi}{\lambda}h_n.
\end{equation}
{The mechanical displacement on the segment $n$, considering a reflective system, is given by $h_n/2$.}

To retrieve the piston and tip-tilt aberrations with the ZELDA signal, we define three estimators, $\varphi_0$, $\varphi_1,$ and $\varphi_2$, following our work with the self-coherent camera-phasing sensor (SCC-PS) in \citet[Eqs.~(17-19) of that paper]{JANIN2016}. Our calculation relies on measurements over an arbitrary square zone $\mathcal{H}$ that is centered on the segment (see Fig.~\ref{fig:zelda_scheme}).
For the piston estimation $\varphi_0$, the signal $I$ is integrated over $\mathcal{H,}$ while for the tip-tilt estimations $\varphi_1$ and $\varphi_2$ the gradient of the signal $I$ is integrated over $\mathcal{H}$ and therefore these estimators are expressed as
\begin{equation}
\label{eq:phi0}
\varphi_0=\iint\limits_\mathcal{H} I(\phi) \mathrm{d}x \mathrm{d}y,
\end{equation}
\begin{equation}
\label{eq:phi1}
\varphi_1=\iint\limits_\mathcal{H} \nabla_x I(\phi) \mathrm{d}x \mathrm{d}y,
\end{equation}
\begin{equation}
\label{eq:phi2}
\varphi_2=\iint\limits_\mathcal{H} \nabla_y I(\phi) \mathrm{d}x \mathrm{d}y,
\end{equation}
where $\nabla_x$ and $\nabla_y$ stand for the gradient along the $\mathbf{x}$ and $\mathbf{y}$ axes.
Using Eqs.~(\ref{eq:signal}) and ~(\ref{eq:elevation}), we derive their theoretical expression:
\begin{equation}
\label{eq:phi0analytic}
\begin{aligned}
\varphi_0\left(p_n,t_n,T_n\right)= &\frac{\sqrt{2}Pb\mathcal{H}^2}{2} \sin\left(\frac{2\pi p_n}{\lambda}-\phi_0 \right) \mathrm{sinc}\left(\frac{\pi \mathcal{H} t_n}{\lambda}\right)
\\
& \times \mathrm{sinc}\left(\frac{\pi \mathcal{H} T_n}{\lambda}\right) + \frac{\mathcal{H}^2}{4}\left(P^2+2b^2\right),
\end{aligned}
\end{equation}
\begin{equation}
\begin{aligned}
\label{eq:phi1analytic}
    \varphi_1\left(p_n,t_n,T_n\right)= &2\sqrt{2} P b \mathcal{H} \cos\left(\frac{2\pi p_n}{\lambda}-\phi_0 \right)
    \\
    &\times \mathrm{sin}\left(\frac{\pi \mathcal{H} t_n}{\lambda}\right)
    \mathrm{sinc}\left(\frac{\pi \mathcal{H} T_n}{\lambda}\right),
\end{aligned}
\end{equation}
\begin{equation}
\begin{aligned}
\label{eq:phi2analytic}
    \varphi_2\left(p_n,t_n,T_n\right)= &2\sqrt{2} P b \mathcal{H} \cos\left(\frac{2\pi p_n}{\lambda}-\phi_0 \right) 
    \\
    &\times \mathrm{sinc}\left(\frac{\pi \mathcal{H} t_n}{\lambda}\right)
    \mathrm{sin}\left(\frac{\pi \mathcal{H} T_n}{\lambda}\right).
\end{aligned}
\end{equation}

\begin{figure*}[!ht]
\centering
\includegraphics[height=0.38\textwidth]{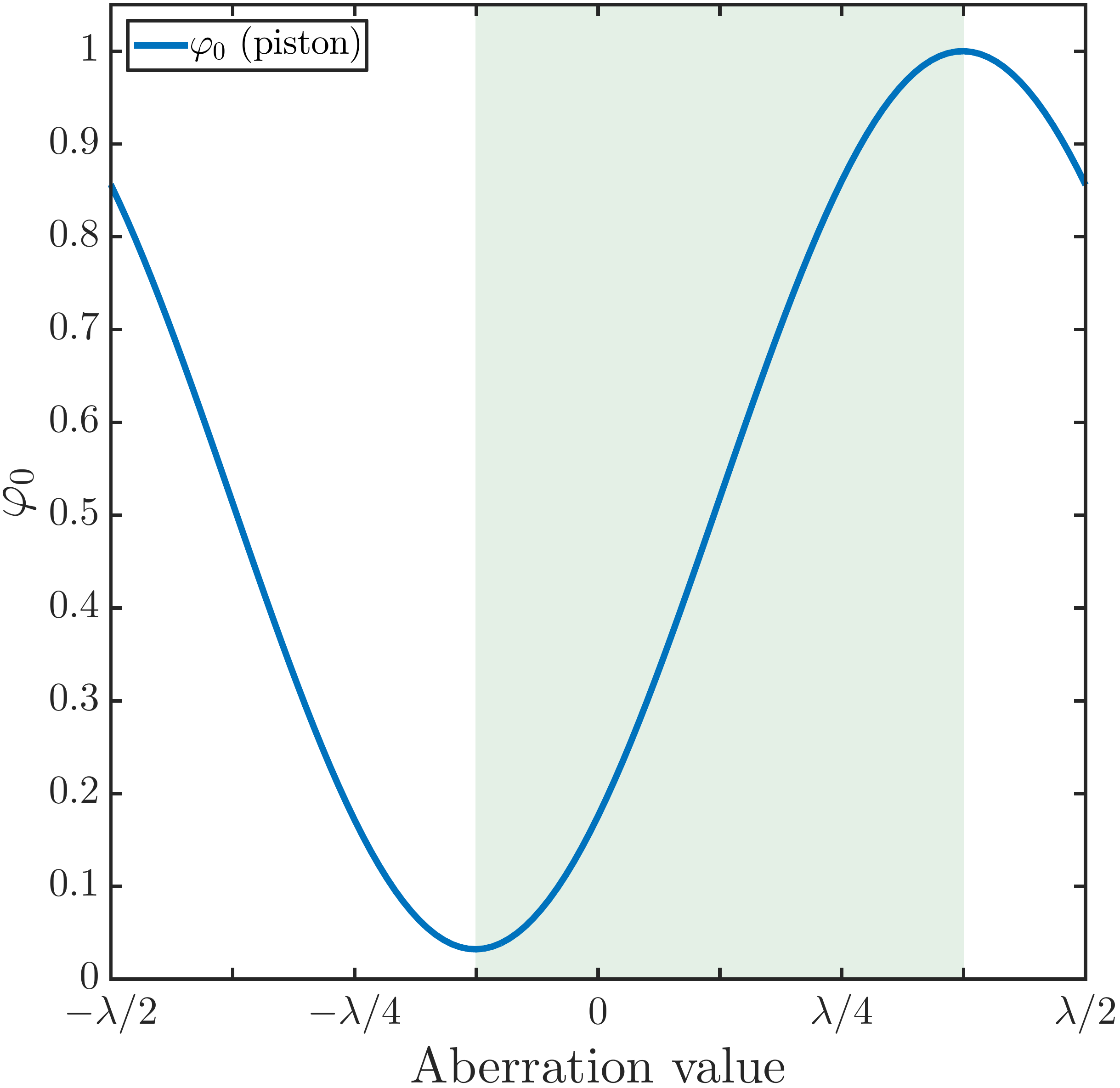}
\hspace{0.08\textwidth}
\includegraphics[height=0.38\textwidth]{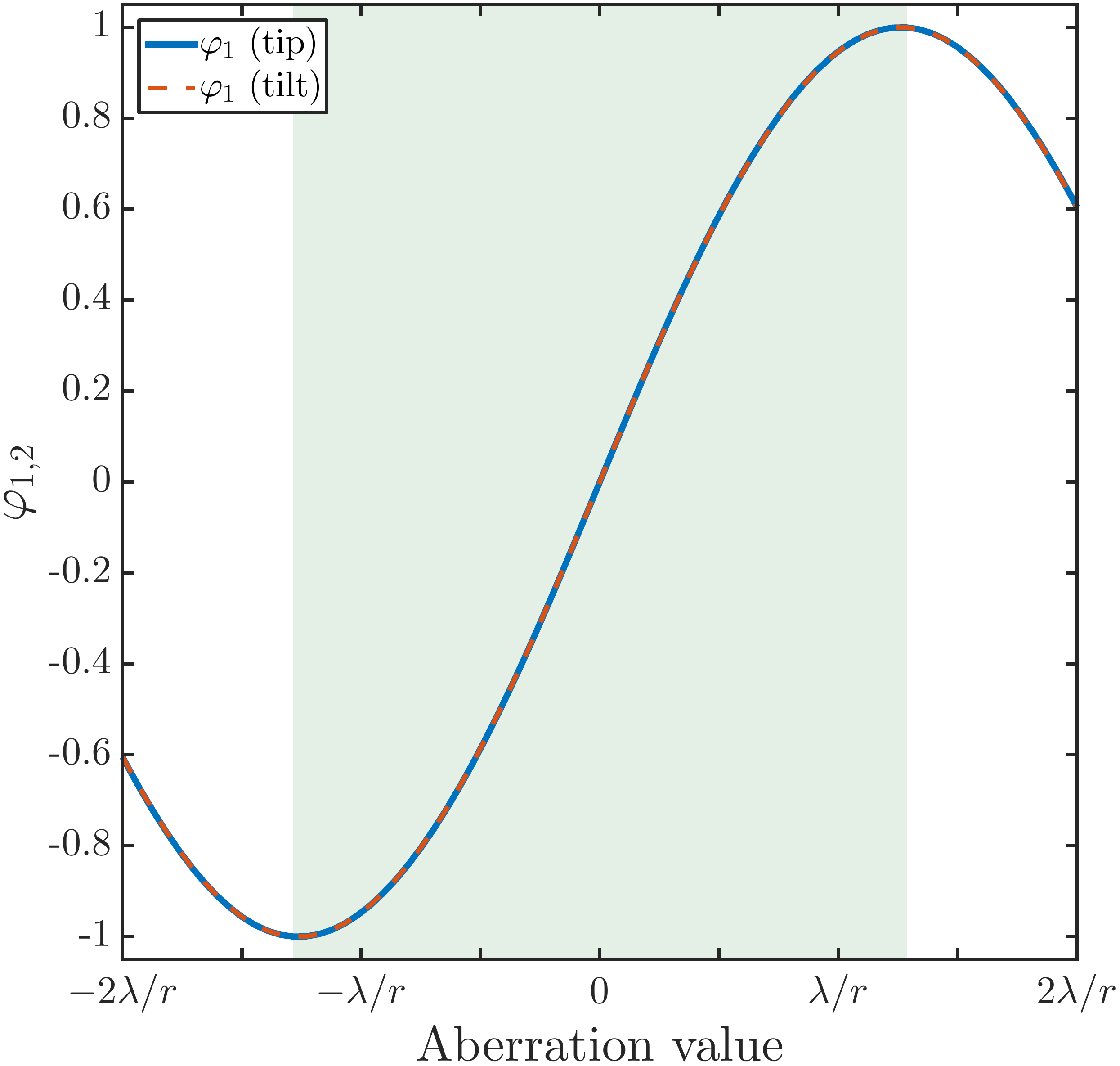}
\caption{Normalized estimators as a function of the introduced aberrations for piston (left) and tip-tilt (right). The green zones represent the capture range for each configuration.}
\label{fig:estimators}
\end{figure*}

\begin{figure*}[!ht]
\centering
\includegraphics[height=0.38\textwidth]{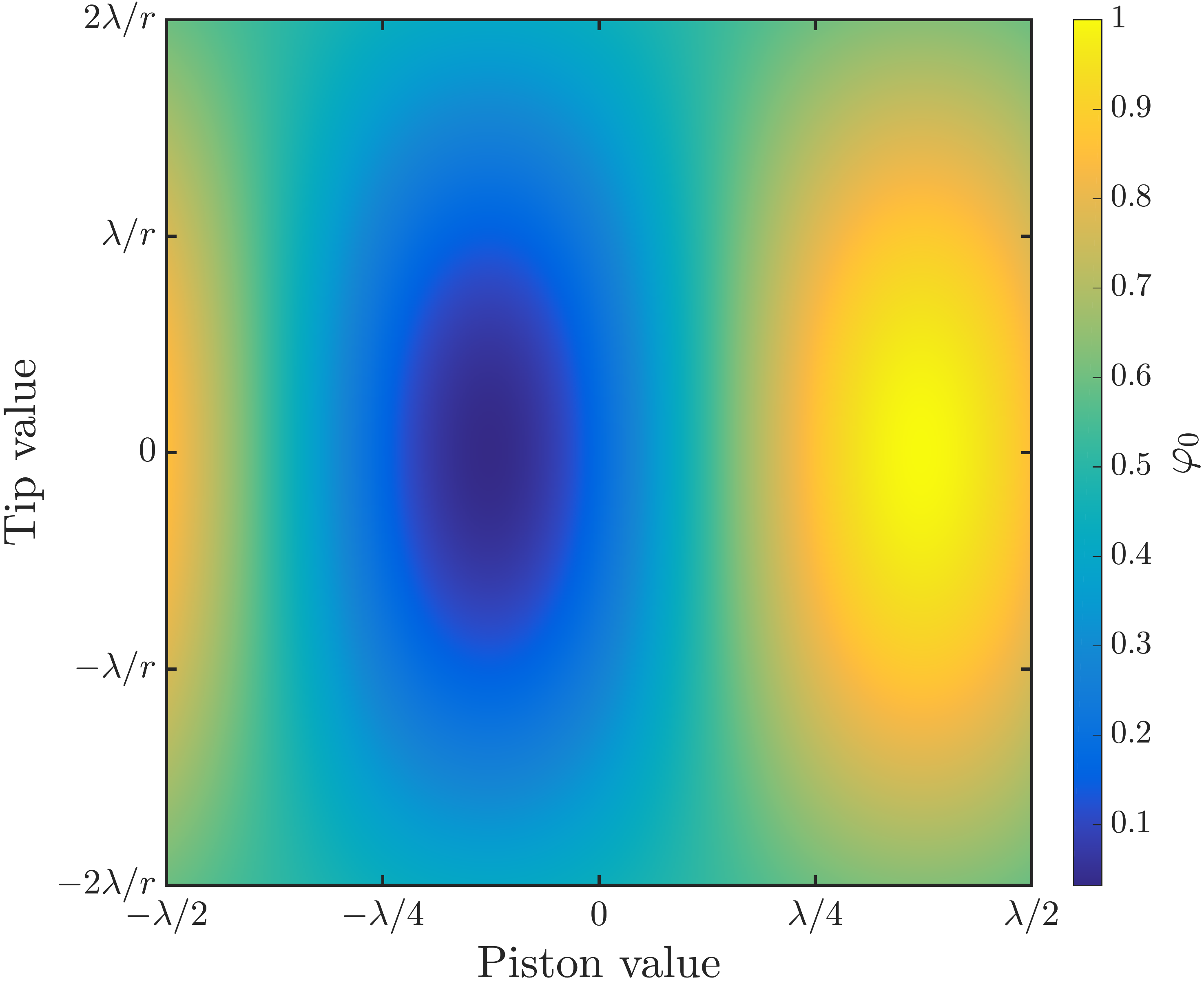}
\hfill
\includegraphics[height=0.38\textwidth]{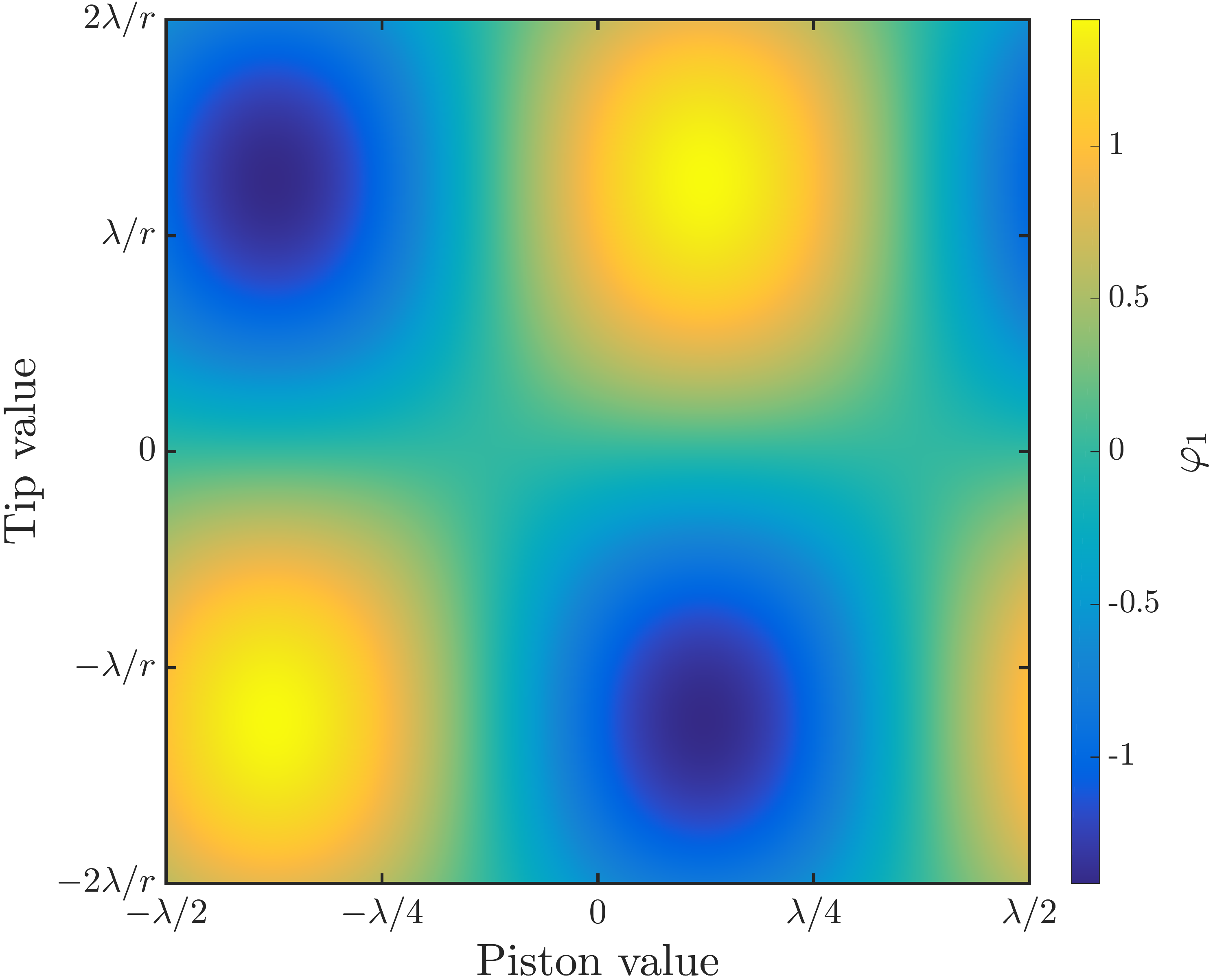}
\caption{Normalized estimators $\varphi_0$ (left) and $\varphi_1$ (right) as a function of the introduced combination of piston and tip.}
\label{fig:estimators2D}
\end{figure*}

\subsection{System response}
\label{subsec:system}

As shown in Eqs.~(\ref{eq:phi0analytic}),~(\ref{eq:phi1analytic}), and~(\ref{eq:phi2analytic}), each estimator depends on piston, tip, and tilt. To investigate this dependence, we assess the behavior of these estimators first with a single aberration and then, with a combination of piston, tip, or tilt on the central segment. {Hereafter, we define $r$ as the radius of the circumscribed circle for a given segment.} In our simulations, we arbitrarily set the $\mathcal{H}$ size to {$0.4r$} that appears as a reasonable trade-off between signal-to-noise ratio (SNR) and sensitivity to pupil shear (see Sec.~\ref{subsec:pupil_misalignment}).

Figure~\ref{fig:estimators} presents the results in the case of a single aberration on the segment. On the left plot, we consider the case of piston only and present $\varphi_0$ response to the introduced piston value. As expected, the estimator $\varphi_0$ is $\lambda$-periodic. This effect represents the well-known $\pi$-ambiguity problem (see, e.g., \citet{VIGAN11}) that limits the reachable capture range in which the measured piston is achieved unambiguously.
The $\varphi_0$ capture range is asymmetric and ranges from $-\lambda/8$ to $3\lambda/8$ as highlighted by the green zone on Fig.~\ref{fig:estimators}. This $\lambda/8$-asymmetry corresponds to the reference phase $\phi_0=\pi/4$ in the ZELDA signal as {shown} in Eq. (\ref{eq:signal}). Such a property will have an impact on the cophasing process for pistons larger than $\lambda/8$ as discussed hereafter in Sec.~\ref{subsec:statistics}.
On the right plot, we present the $\varphi_1$ and $\varphi_2$ responses to the introduced tip and tilt values.
Both estimators are periodic and symmetric with a capture range that is related to $\mathcal{H}$ and extends from -$5\lambda/4r$ to $5\lambda/4r$. In the absence of piston, large segment tip/tilt can be estimated with our sensor signal.  

Figure~\ref{fig:estimators2D} presents the same rationale but with combined piston, tip, and tilt. The left plot represents $\varphi_0$ for a combined segment piston and tip. The correlation between $\varphi_0$ and the two introduced aberrations is clearly visible. As the tip increases in absolute value, the $\varphi_0$ estimation decreases following a cardinal sine shape, leading to an underestimation of the piston. Even though not critical, this dependence has an influence on the cophasing process and is studied in Sec.~\ref{subsec:statistics}.
A correlation is also observed on Fig.~\ref{fig:estimators2D} (right) when measuring $\varphi_1$ in the presence of piston and tip. The same results are obtained if we measure $\varphi_2$ for a combination of piston and tilt, as suggested by Eqs.~(\ref{eq:phi1analytic}) and (\ref{eq:phi2analytic}).
As tip varies, the $\varphi_1$ estimation oscillates in a sinusoidal shape. The tip estimation is therefore strongly affected by the presence of piston. A sign change of the estimator is even observed for large values of piston, leading to a misinterpretation of the segment orientation. We also note that $\varphi_1$ can be larger than $1$. This particular behavior comes from the normalization of $\varphi_1$ by its peak value at null values of piston and tilt. When the piston is equal to $\lambda/8$, the normalized $\varphi_1$ value reaches $\sqrt{2}$. Both effects can prove critical in aberration measurements and impact the segment cophasing process. We discuss the operating mode in Sec.~\ref{subsec:statistics}.


\begin{figure*}[!ht]
\centering
\includegraphics[height=0.38\textwidth]{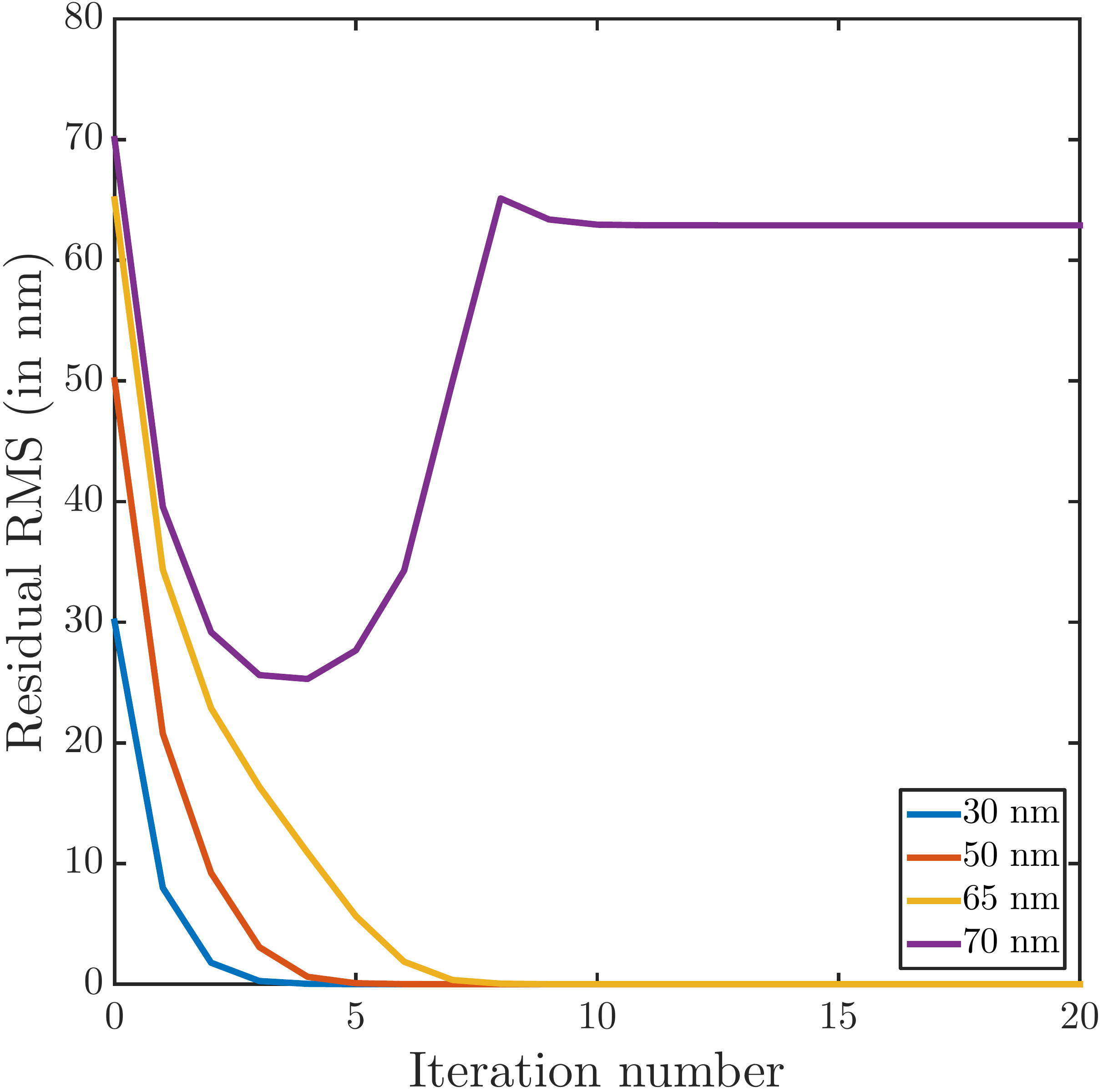}
\hspace{0.08\textwidth}
\includegraphics[height=0.38\textwidth]{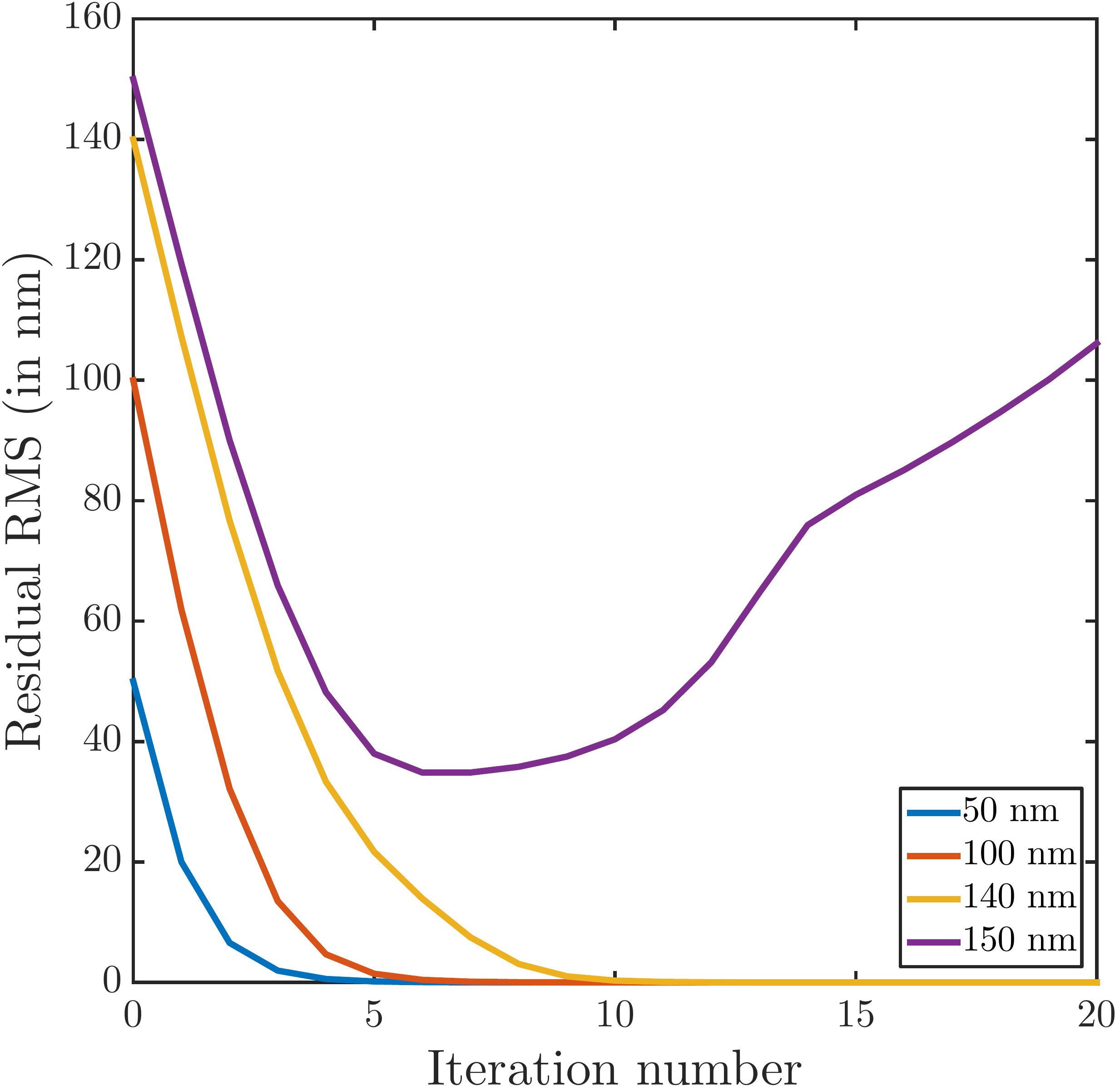}
\caption{Iterative evolution of the residual RMS wavefront error for piston only (left) and for combined piston and tip-tilt (right) with different initial conditions.}
\label{fig:closeloop}
\end{figure*}

\section{System calibration}
\label{sec:numerical}

\subsection{Numerical assumptions}

Numerical simulations are performed with a pupil made of 91 hexagonal segments that are distributed over five concentric rings as shown in Fig.~\ref{fig:zelda_scheme}. Each segment is sampled with 100 pixels from corner to corner {(i.e., $r=50$~pixels)}. In this study, no central obstruction nor spiders are taken into account as they have no incidence on the ZELDA measurement. To ensure a fine sampling within the focal plane mask and a fast computation, the ZELDA signal is calculated using the semi-analytical method based on the Matrix Fourier Transform {(MFT)} described by \citet[]{SOUMMER07}. For the phase mask, we set $d$ to $1.06\lambda/D$ and sample it with $500$~pixels.
{We consider a source in monochromatic light at wavelength $\lambda=600$~nm.}

\subsection{Calibration and properties}

The calibration process of the system is based on the traditional adaptive optics scheme. A calibration matrix is built by actioning successively and independently each segment in piston, tip, and tilt, and by measuring $\varphi_0$, $\varphi_1,$ and $\varphi_2$. The amplitude of the displacements is chosen within the linear regime of the sensor. In practice, the poke values are set to $1/10$ of the capture range for each estimators, that is, $\lambda/20$ for piston and $\lambda/4r$ for tip and tilt.
The calibration process is done using a perfectly aligned and {flattened} system. At this stage, no camera noise or photon noise are taken into account.
The calibration matrix is square and has dominant diagonal terms. However, some patterns are visible outside the diagonal and are due to the correlation between the three estimators.
Finally, piston, tip, and tilt errors for each segment are retrieved by solving the linear system of equations by means of the inverted calibration matrix and are applied to the segments.
{In any on-sky application, the calibration matrix is conventionally measured using either an internal calibration source or a stellar target. A pseudo-synthetic calibration matrix can also be built by experimentally measuring the estimators on a single segment and replicating them in all the other segments in the matrix structure deduced from simulations \citep[e.g., ][]{YAITSKOVA06}.}
While straightforward when {looking at} a reasonable number of segments, the calibration matrix is avoidable by using an alternative method based on the ZELDA signal normalization.

The system evolves in a closed-loop architecture and the convergence quality is assessed by using the residual root mean squared (RMS) wavefront error $\sigma_{pupil}$ over the entire pupil. Its analytical expression is given by
\begin{equation}
\begin{aligned}
\label{eq:STD}
\sigma_{pupil}= \sqrt{s_p^{\,2} + \frac{5}{24} r^2\left( s_t^{\,2} + s_T^{\,2} + \mathbb{E}\left[\mathbf{t}\right]^{\,2} + \mathbb{E}\left[\mathbf{T}\right]^{\,2} \right)},
\end{aligned}
\end{equation}

\noindent where $s_p$, $s_t$ , and $s_T$ stand for the piston, tip, and tilt samples standard deviation, and $\mathbb{E}[...]$ denotes the mathematical expectation. A detailed development of Eq. (\ref{eq:STD}) is given in Appendix~\ref{app:std}.

\section{Performance and discussions}
\label{sec:results}
We now present the results for the complete computation of a full segmented pupil in contrast to Sect. \ref{subsec:system} in which the response of an individual segment has been characterized. To assess the performance of the system, we proceed to the following tests: (1) a closed-loop convergence evaluation; (2) a statistical analysis of the converging process; (3) a sensitivity analysis of hardware misalignment combined with probabilistic insights; and finally (4) a sky coverage evaluation.

\begin{figure*}[!ht]
\centering
\includegraphics[height=0.38\textwidth]{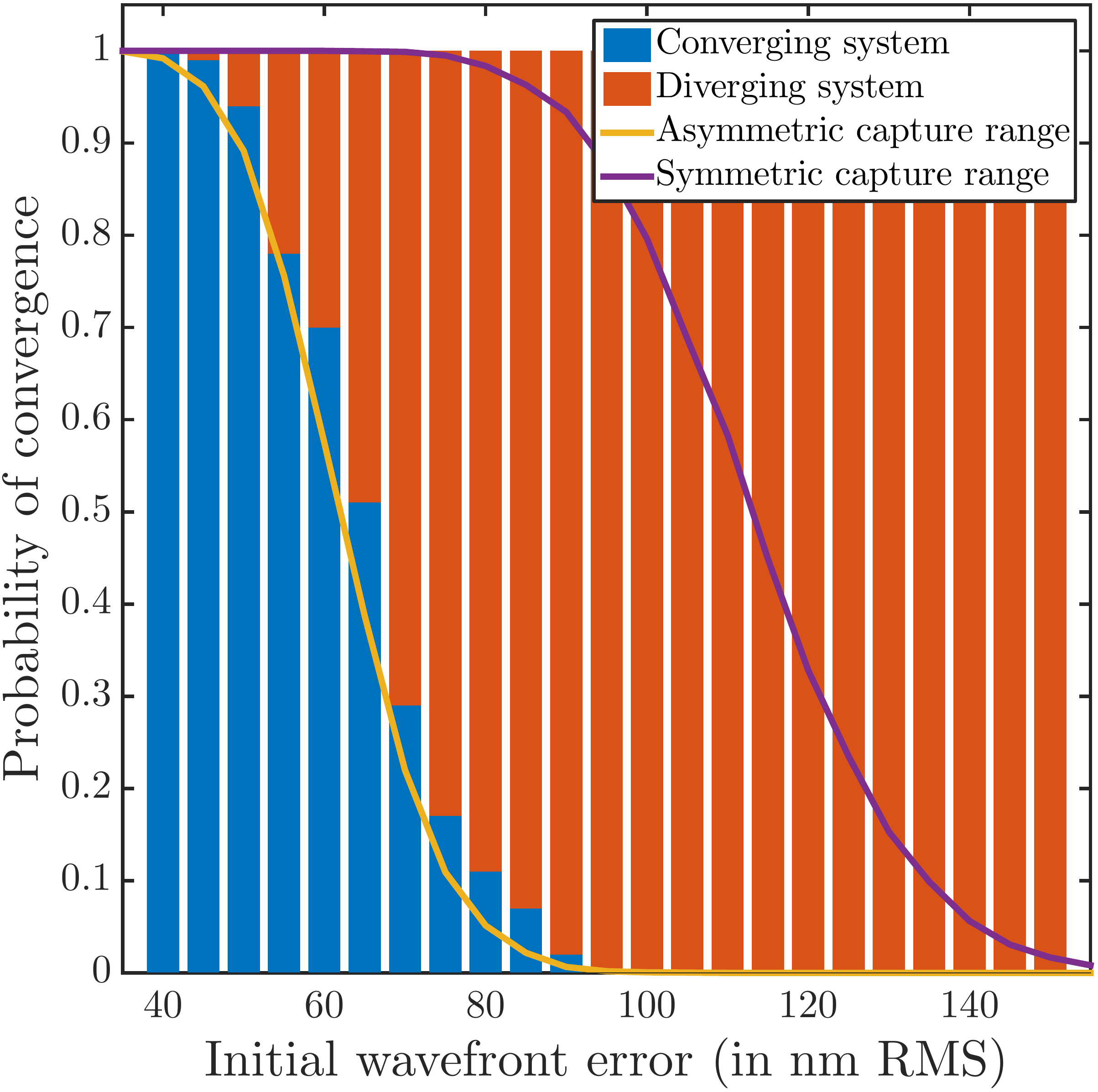}
\hspace{0.08\textwidth}
\includegraphics[height=0.38\textwidth]{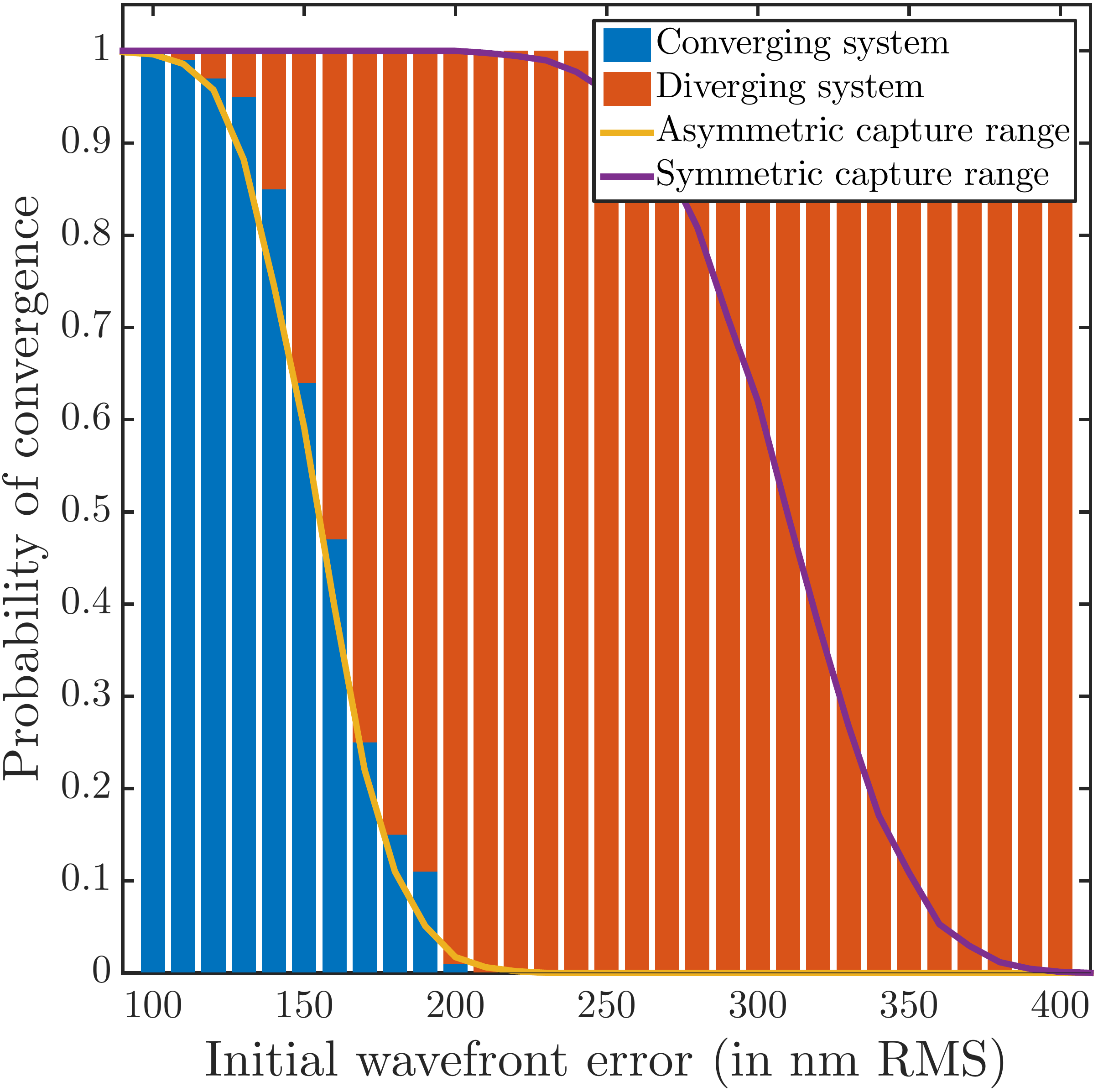}
\caption{Probability of convergence as a function of the initial RMS wavefront error for piston only (left) and piston and tip-tilt (right).}
\label{fig:statistics}
\end{figure*}

\subsection{Closed-loop accuracy}
\label{sec:closed-loop}

The results of the closed-loop operation on the phasing process for different initial wavefront errors (expressed in nanometer RMS) are presented in Fig.~\ref{fig:closeloop}. 
We differentiate between the results for the piston error only (left) and for the  combined piston, tip, and tilt errors (right), where each case uses its specific calibration matrix that includes the considered aberrations only.
\if
A set of aberrations $[\mathbf{p_0},\mathbf{t_0},\mathbf{T_0}]$ with a unity variance is created by normally distributing each $p_n$, $t_n$, and $T_n$ with the standard deviations $\sigma_p$, $\sigma_t$ and $\sigma_T$ respectively. The contribution of each aberration is set to the same fraction $\beta$ of the capture range using the following relations:
\begin{equation}
    \begin{array}{lcr}
        \sigma_p&=&\beta\times(\lambda/2),
        \\
        \sigma_t&=&\beta\times (5\lambda/2\mathbf{r}),
        \\
        \sigma_T&=&\beta\times (5\lambda/2\mathbf{r}).
    \end{array}
\end{equation}
\fi
{A set of reference aberrations $[\mathbf{p_0},\mathbf{t_0},\mathbf{T_0}]$ is created by drawing each $p_n$, $t_n$, and $T_n$ from a Gaussian distribution with zero mean and $\sigma_p$, $\sigma_t$ , and $\sigma_T$ standard deviations. In order to insure an equal contribution of each aberration, the standard deviations are set according to their respective capture range as
\begin{equation}
    \begin{array}{lcc}
        \sigma_p&=&\lambda/2,
        \\
        \sigma_t&=& 5\lambda/2r,
        \\
        \sigma_T&=& 5\lambda/2r.
    \end{array}
\end{equation}
}
A practical set of aberrations are then scaled using a gain factor $\alpha$ as $\alpha \times [\mathbf{p_0},\mathbf{t_0},\mathbf{T_0}]$ to produce different levels of initial RMS wavefront errors.
{
The typical cophasing process occurs as follows:
\begin{enumerate}
    \item Set $\alpha$ such that $\sigma_{pupil}(\alpha\times[p_0,t_0,T_0])$ is equal to the initial wavefront error we want to start with.
    \item Compute a ZELDA image using the MFT.
    \item Calculate $\varphi_0$, $\varphi_1$, $\varphi_2$ and convert to piston, tip, and tilt using the calibration matrix.
    \item Calculate the residual piston, tip, and tilt expected after moving the segments.
    \item Repeat steps 2-4 until convergence is reached or for a given number of iterations.
\end{enumerate}
}

Figure \ref{fig:closeloop} exhibits two phasing regimes: (1) the case for which the segments are being phased, that is, when the zero residual error is achieved, and (2) the case where one or more segments are shifted by an integer number of the wavelength (with the increase in the initial wavefront error, some segments are left outside the capture range), leading to a non-phased mirror. The last case (2) is known as the \textit{$\pi$-ambiguity} problem, common to any phasing sensor operating in the monochromatic regime. 

The results of the phasing process for the piston error only (Fig. \ref{fig:closeloop}, left) demonstrate that the system converges to zero error residual for initial aberration values that are smaller than $65$~nm RMS. {This capture range} is below what is generally reachable with state-of-the-art phasing sensors. 
This {reduction in dynamic range} is a direct effect of the asymmetric piston capture range that is specific to ZELDA as described in Sec. \ref{subsec:analytical} and explained in Sec. \ref{subsec:statistics}.
For larger aberrations, the system converges to a stable but non-phased state (green curve in Fig. \ref{fig:closeloop}, left).
When the piston, tip, and tilt errors are combined (Fig.~\ref{fig:closeloop}, right), the observations are roughly identical, but the threshold is higher and equal to $140$~nm RMS {(for information, $\sigma_p/\sigma_{pupil}\sim0.3$)}. Nonetheless, Fig.~\ref{fig:closeloop} (right, purple curve) shows that when the system converges to a non-phased mirror, the system is unstable. While this instability is not observed for piston-only error, it here originates from the correlation between the three estimators $\varphi_0$, $\varphi_1$, and $\varphi_2$. When aberration amplitudes become too large, a rapid amplification of the estimation error due to cross-terms occurs and inevitably conveys the system to a chaotic regime.   

\subsection{Probability of convergence}
\label{subsec:statistics}

In this section, we investigate the probabilistic behavior of the phasing process with ZELDA. The objective is to propose from a statistical point of view a predictive model of convergence as a function of the initial aberration error value. 
In the following, the starting wavefront errors correspond to a fraction of the capture range as already explained in Sec. \ref{sec:closed-loop}.
Piston error only (1) and piston combined with tip-tilt (2) are addressed separately for the sake of clarity.

In case (1), both empirical and semi-analytical approaches are carried out.
The empirical method is performed by considering the outcome of 100 samples of phasing attempt processes. The approximated probability of convergence for a given variance is then
\begin{equation}
\mathbbm{P}_{E}(\sigma_{pupil}^{\,2})=n_{conv}/100,
\end{equation}
where $n_{conv}$ is the number of positive outcomes, referring to the situation for which the phasing process achieves a phased and stable mirror state, that is, the phasing process converges. Alternatively, negative outcomes correspond to the case of the diverging phasing process.  
Figure~\ref{fig:statistics} (left) presents the occurrences for converging cases (blue bars) and diverging cases (red bars). We note that $\mathbbm{P}_1=0.5$ for $\sigma_{pupil} \simeq 65$~nm RMS.
\\
The semi-analytical approach is based on a 10000-sample Monte Carlo integration used to infer the probability $\mathbbm{P}_{SA}$ for each segment to be within the given range $[a,b]$ around its mean position. This probability is expressed as

\begin{equation}
\begin{aligned}
\label{eq:proba_range}
\mathbbm{P}_{SA}(\sigma_{pupil}^{\,2}, a, b)= \int_{\left(\mathbbm{R}\right)^{N}} \mathbbm{1} \left\{ \bigcap\limits_{n=1}^{N} \left(a < p_n - \mathbb{E}\left[\mathbf{p}\right]  < b \right) \right\}
\\
\times \prod_{n=1}^{N}  f_P(p_n \mid 0,\sigma_p^{\,2}) \left( \mathrm{d} p_1 \cdots \mathrm{d} p_n \cdots \mathrm{d} p_{N} \right),
\end{aligned}
\end{equation}
where $\mathbbm{1}$ is the indicator function. Since the introduced pistons are randomly issued from a zero-mean normal distribution, $f_P(p \mid 0,\sigma_p^{\,2})$ is the probability density of the Gaussian distribution $\mathcal{N}(0,\sigma_p^{\,2})$.
Based on Fig.~\ref{fig:estimators}, we set $a$ and $b$ to $-\lambda/4$ and $3\lambda/4$ respectively, that strictly delimits the domain outside which the system corrects for in the wrong direction.
The result of the Monte Carlo integration corresponds to the yellow curve presented on Fig.~\ref{fig:statistics} (left), and it fits well with the data from the empirical method. 
This provides confidence in the initial assumption that a system converges when all the individual pistons over the pupil are within the range $\left[\mathbb{E}\left[\mathbf{p}\right]-\lambda/4,\mathbb{E}\left[\mathbf{p}\right]+3\lambda/4\right]$.
\begin{figure*}[!ht]
\centering
\includegraphics[height=0.38\textwidth]{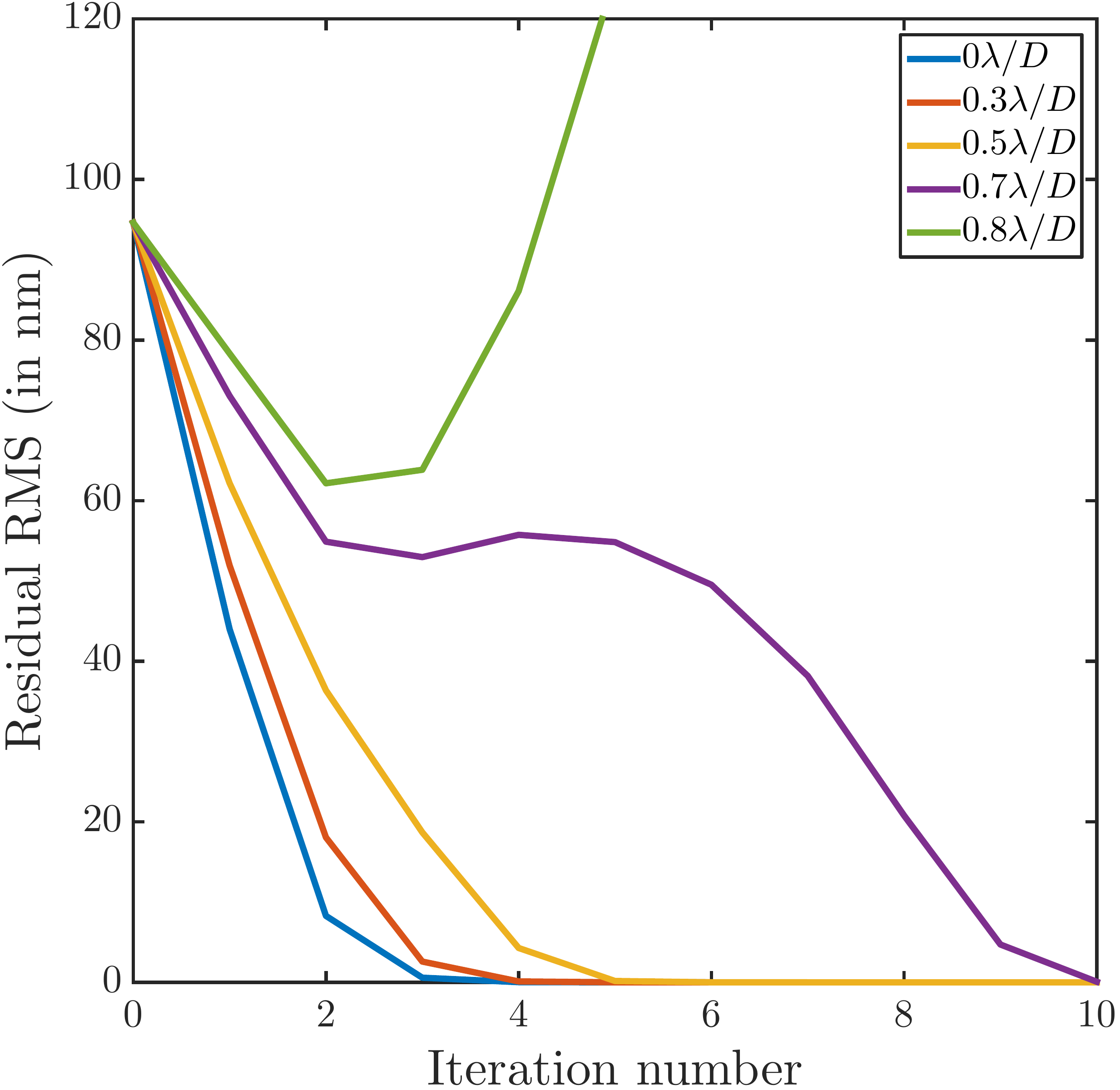}
\hspace{0.08\textwidth}
\includegraphics[height=0.38\textwidth]{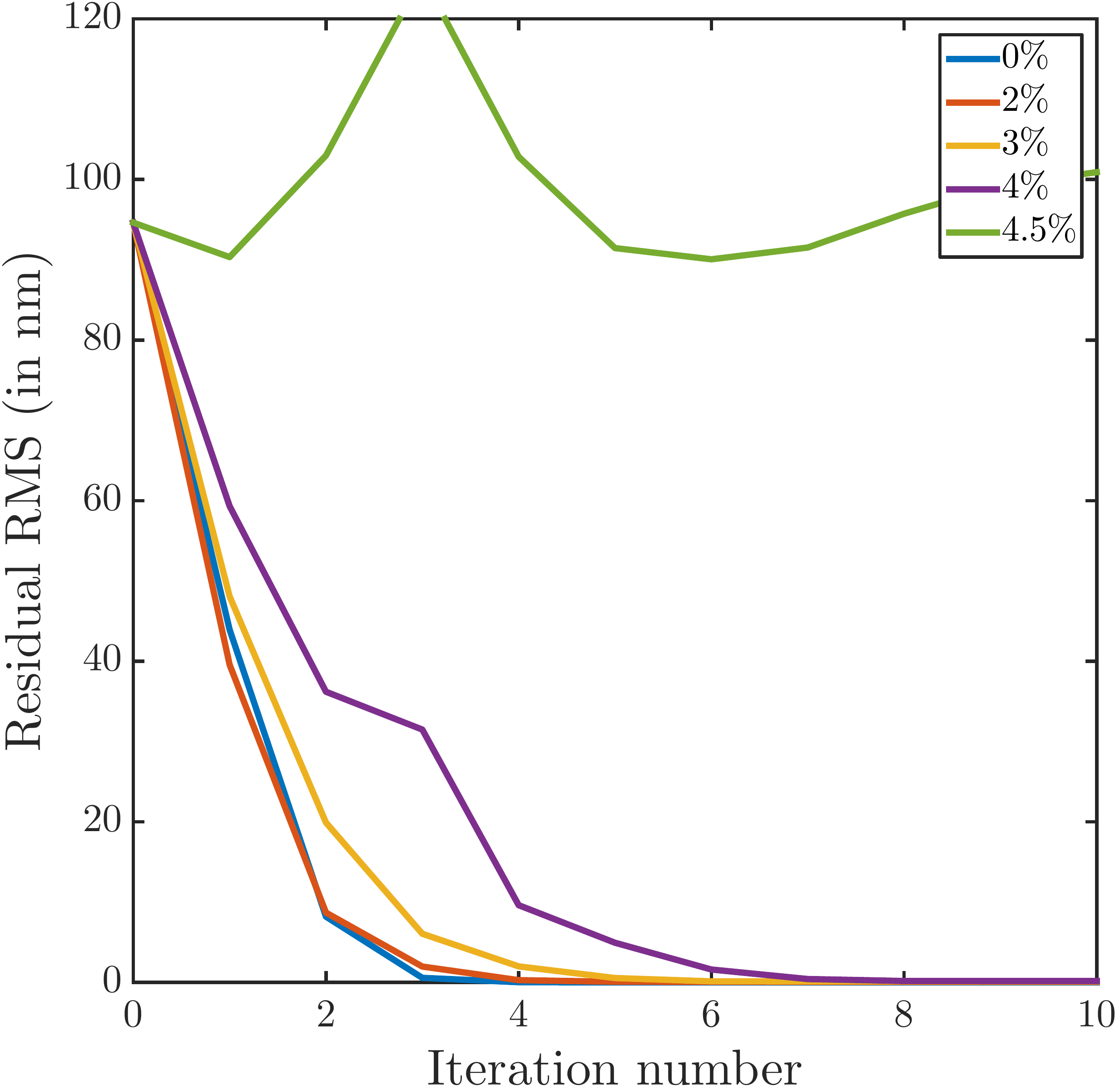}
\caption{Iterative evolution of the residual RMS wavefront error as a function of the focal mask (left) and pupil (right) displacements.}
\label{fig:misalignment}
\end{figure*}
Actually, unlike ZELDA, a phasing system with a symmetric capture range will converge for larger initial aberrations: shifting only the current boundaries to $a=-\lambda/2$ and $b=\lambda/2$ transforms the yellow curve into the purple one (see in Fig.~\ref{fig:statistics} (left)). As mentioned in Sec.~\ref{sec:closed-loop}, a change in the phasing capture range inevitably translates into a change in the convergence dynamic.

In case (2), two complementary approaches are also used. The same empirical method as previously defined is performed by producing 100 samples of system iterations and evaluating the outcomes to derive $\mathbb{P}_E$. Figure~\ref{fig:statistics} (right) presents the results, where the blue bars indicate the converging cases and the red one the diverging cases.
\\
The second method is different from the semi-analytical approach that has been previously used for the case of piston only. With the correlation between piston, tip, and tilt estimators, elementary rules of thumb on the initial conditions cannot be settled to frame the convergence outcomes. 
To overcome this issue, we use the theoretical expressions of the estimators shown in Eqs.~(\ref{eq:phi0analytic}-\ref{eq:phi2analytic}) to emulate the converging process where only the diagonal terms of the calibration matrix are kept.
The results are represented by the yellow curve in Fig.~\ref{fig:statistics} (right) and match the previous data.
A slight offset, however, indicates that the method tends to underestimate the number of converging systems. This effect is also distinguishable on the piston-only configuration. This underestimation reflects the shortcoming of a binary condition to predict the outcome of the phasing process. In an actual system, while few segments might initially be out of the capture range, and since the estimators are correlated, it might occur against the odds that the amplitude of misalignment of these segments meets the boundaries of the capture range, and finally succeeds in converging to zero error residuals. 
Finally, by analogy with case (1), a phasing system with a symmetric capture range will converge for larger initial aberrations (Fig.~\ref{fig:statistics}, right, purple curve).

The models defined in this section allow one to infer the probability of convergence for a given initial wavefront error. Although the presented results have been obtained with a pupil made of 91 segments, the models can account for different configurations.
The limit between the cophasing regimes can be clearly set based on a given probability threshold. Fixing this probability to $P=95\%$, one can conclude that for an asymmetric sensor the limit between fine and coarse phasing is 120~nm, while for a symmetric sensor this limit is 250~nm for piston and tip-tilt cases.
These algorithms can be used for preliminary studies of a system to conjecture the needs in terms of cophasing accuracy and capture range.

\subsection{Sensitivity to misalignment}

The sensitivity of the cophasing system regarding hardware alignment is studied by investigating both the FPM and the pupil misalignment. For the sake of similarity with the JWST, we consider a pupil made of $19$ segments. 
A specific vector of initial aberrations $[\mathbf{p_1},\mathbf{t_1},\mathbf{T_1}]$ is created (see Sec. \ref{sec:closed-loop}). 
Because the phasing process is subject to statistically driven behaviors, these aspects are addressed in Sec.~\ref{subsec:stat_mis}. 

\subsubsection{Focal plane mask misalignment}
\label{subsec:foc_misalignment}

To determine the effect of the FPM misalignment, we evaluate the convergence of the system
and record the residual wavefront error at each step for several FPM displacements. This process is repeated for various sets of aberration over the pupil. 
Figure~\ref{fig:misalignment} (left) presents the results for a pupil exhibiting an initial aberration of $100$~nm RMS. This value is chosen for the system to have a probability to converge $P\sim1$.  We note that the residual global tip-tilt over the pupil, originating from the system compensation for the FPM misalignment, is numerically removed from the data (the telescope active optics correct for the global tip-tilt and defocus). 
Figure~\ref{fig:misalignment} (left) shows that the system can accommodate a FPM misalignment up to $0.7 \lambda/D$. Since levels of mask positioning can be guaranteed to a level of $10^{-2}$ to $10^{-3}$ $\lambda/D$ using a low-order wavefront sensor \citep[e.g., ][]{MAS12}, the sensitivity of our closed-loop system to FPM misalignment is negligible. 

\begin{figure}[!ht]
\centering
\includegraphics[width=0.38\textwidth]{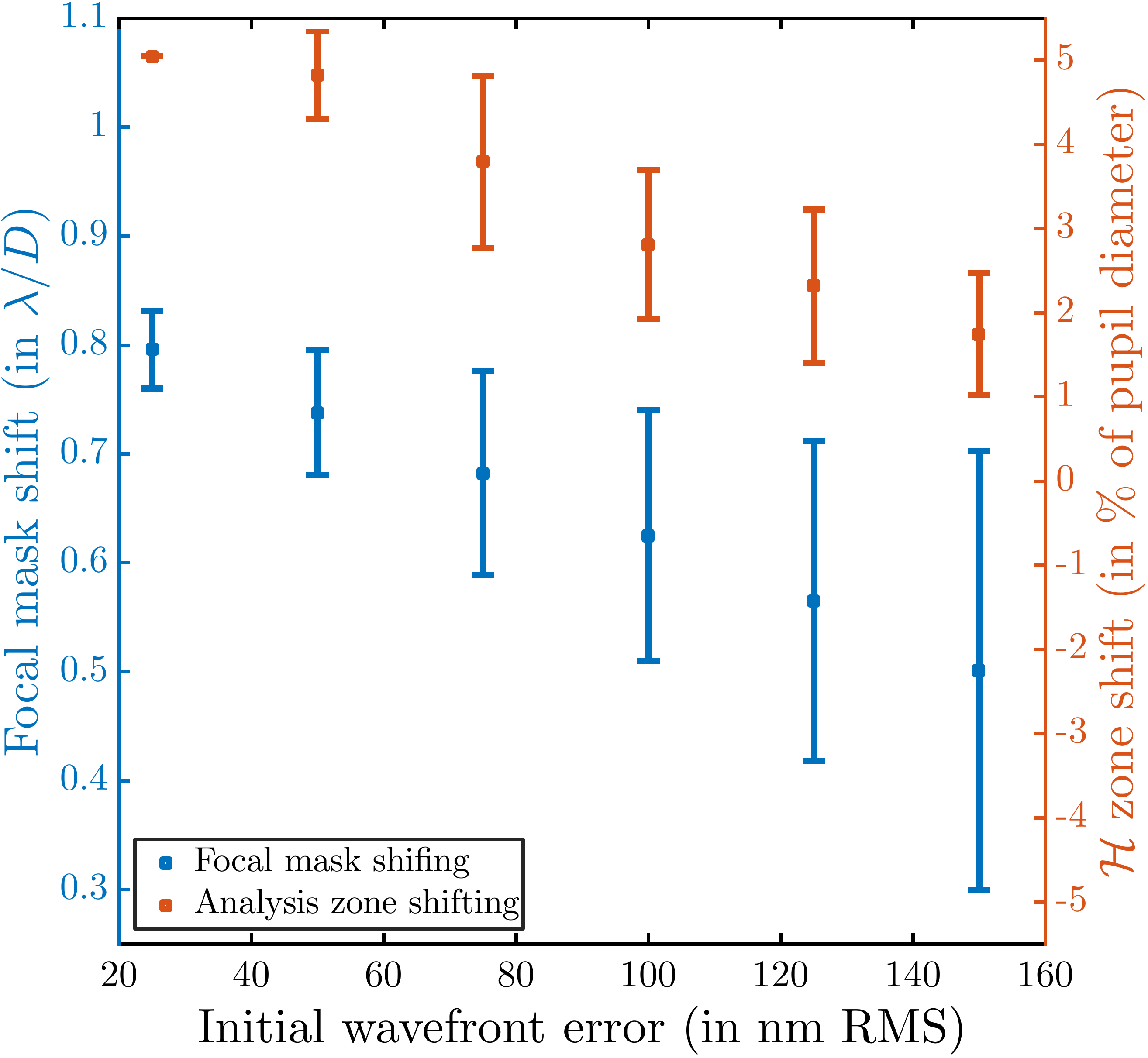}
\caption{Sensitivity to focal mask (blue) and pupil (red) displacements as a function of the initial aberration wavefront error. Error bars represent $1\sigma$ dispersion.}
\label{fig:stat_displacement}
\end{figure}

\subsubsection{Pupil misalignment}
\label{subsec:pupil_misalignment}

In real observing conditions, the position of the pupil {may be shifted by} a significant fraction of its diameter \citep[$3$-$4\%$ for the JWST, ][]{BOS2011}. This can possibly affect the phasing process as the calibration matrix is built on a perfectly centered system.
We study this effect by numerically shifting the analysis zone $\mathcal{H}$ and processing the closed-loop operation. 
\if
{À insérer ici ? -> The zone are moved in the same direction which is consistent with a global displacement of the pupil. Segmented telescopes can however be affected by deformations within their pupil. Assuming that this effect does not significantly add piston and tip-tilt aberrations, the results should be equivalent because of the statistical independence between the segments.}
\fi
Figure~\ref{fig:misalignment} (right) presents the results of the convergence for several displacements. As for the FPM misalignment analysis, the initial wavefront error is about $100$~nm RMS (the global tip-tilt is also removed from the data). For displacements as large as $4\%$ of the pupil diameter, {corresponding in our simulations to $17.5\%$ of the segment diameter}, the system converges. This indicates a low sensitivity of the sensor to the pupil shear. We propose a detailed statistical study of these aspects in the following.

\subsubsection{Statistical behavior of the misalignment}
\label{subsec:stat_mis}

We estimate the limit in FPM and pupil displacements which the system can accommodate for different initial RMS wavefront error.
Figure~\ref{fig:stat_displacement} presents the results for the FPM (blue points) and the pupil (red points) displacements. The mean value has been computed for $100$ realisations. The error bars represent the $1\sigma$ standard deviation.
As expected for both FPM and pupil analysis, the larger the initial wavefront error the lower the tolerance.
Nevertheless, a difference in the standard deviation of the FPM and pupil results is observable. 
While the standard deviation is roughly constant with the FPM displacement, it increases significantly with the pupil misalignment. Pupil and FPM displacements therefore impact on the system differently.

\begin{figure}[!ht]
\centering
\includegraphics[width=0.38\textwidth]{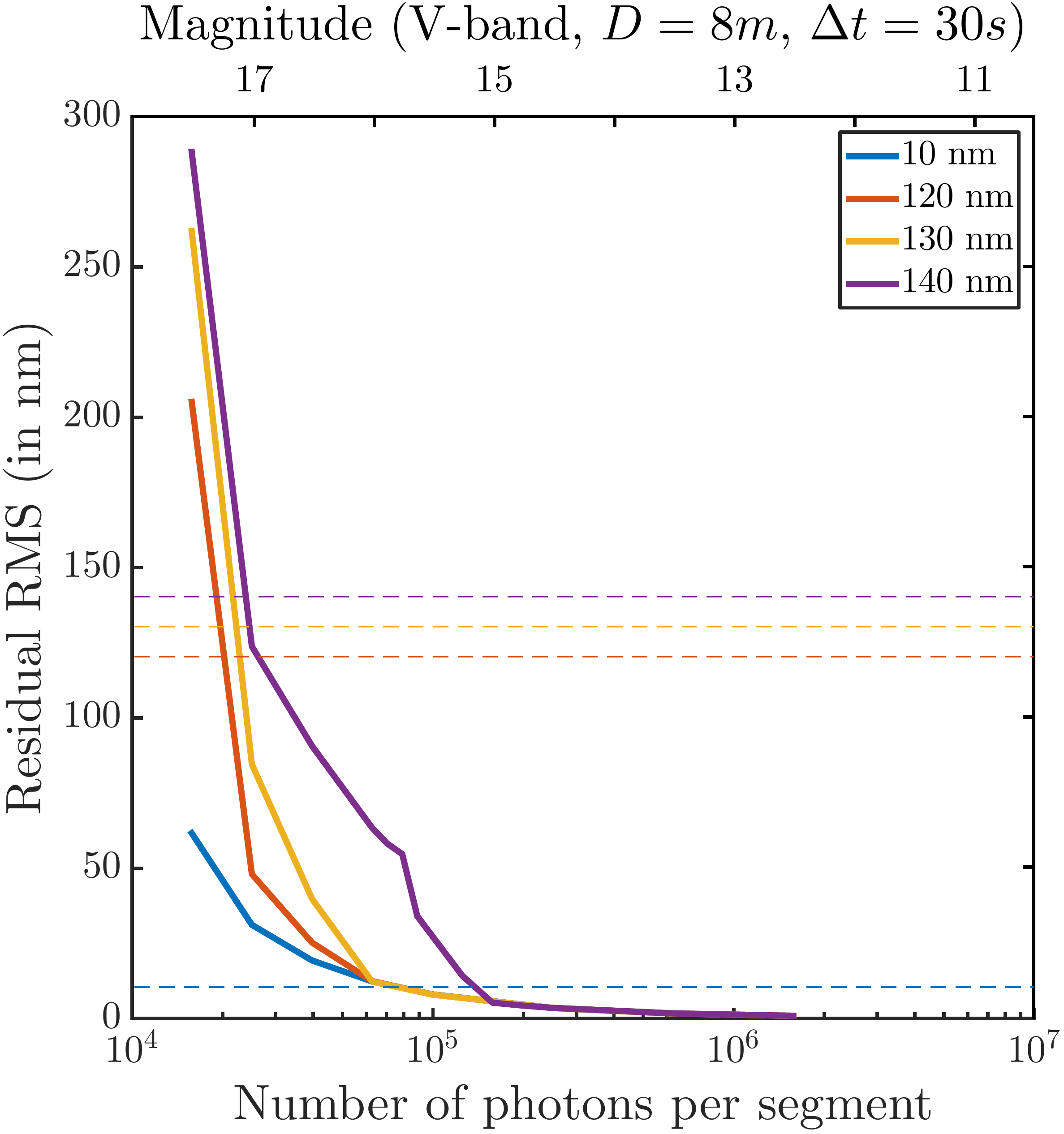}
\caption{Residual RMS wavefront error over the pupil after 20 iterations as a function of the number of photons per segment for different initial aberrations. The dashed lines represent the initial aberration levels.}
\label{fig:magnitude}
\end{figure} 

Analytical development of the estimators shows that the pupil displacement affects the measurement on each segment by introducing a systematic error proportional to  the displacement. All the segments are impacted   alike and have the same statistical weight when considering their contribution to initiate the phasing process divergence.
The effect is different when considering the contribution of a FPM displacement.
Numerical simulations show that the system converges towards a configuration with a global tip-tilt over the pupil (compensating for the FPM displacement), meaning that the final position of each segment should {converge to}
\begin{equation}
[p_n,t_n,T_n]=[p_n^*,t^*,T^*],
\end{equation}
where $t^*$ and $T^*$ stand for the global tip and tilt respectively, and $p_n^*$ the residual piston associated to the segment $n$ and subject to $t^*$ and $T^*$. The expression for the piston on the $n^{th}$ segment is
\begin{equation}
\label{eq:piston_star}
p_n^*=\frac{r}{2}\left[3k_n t^* + \sqrt{3}l_n T^*\right] \quad \forall(k_n,l_n) \in \mathbbm{Z},
\end{equation}
where $r$ denotes the segment radius; $k_n$ and $l_n$ are specific to each segment and their values depend on the individual segment position in the pupil.
The FPM displacement contribution affects the piston distribution non-uniformly in the pupil, and thus alters the estimators differently on each segment. Outer ring segments are more impacted and are thus predisposed to initiate the phasing process divergence well before the inner ring ones.

To summarize, while the contribution of the pupil displacement is local, the FPM displacement contribution is global. Thus the number of segments capable of initiating the process of divergence is different. While for a pupil displacement borderline segments are equiprobable, for a FPM displacement outer ring segments are over-weighted. This can explain the widening of the standard deviation observed in Fig. \ref{fig:stat_displacement}.

\subsection{{Stellar magnitude sensitivity}}
\label{subsec:magnitude}

As for hardware misalignment, the sensitivity to stellar magnitude is critical to evaluate the capacity of the sensor to operate under realistic conditions. 
The analysis is performed assuming {an} $8$~m telescope observing in V-band (spectral bandwidth $\Delta\lambda =90$~nm) for $30$~s integration time. The pupil is composed of $91$~segments. 
In these simulations, we assume photon noise, readout noise ($1~\mathrm{e}^{-}$ RMS), and dark current ($1~\mathrm{e}^-\mathrm{/s}$). 
We use the residual RMS wavefront error after $20$ iterations as a metric to determine the quality of the convergence. To smooth out the random behavior of the process, the median of $50$ different sets of aberrations is considered.
Figure~\ref{fig:magnitude} presents the results and as expected, the {sensitivity to stellar magnitude} depends on the initial RMS wavefront error. Simulations show that ZELDA {has a sensitivity to stellar magnitude} that is consistent with other state-of-the-art phasing sensors \citep[e.g., ][]{PINNA08, GONTE2009, SURDEJ10}. {Assuming a V magnitude of $15$, ZELDA provides a residual wavefront error smaller than $5$~nm RMS for an initial wavefront error as large as $140$~nm RMS}.

\section{Conclusion}
\label{sec:conclusion}
In this paper, we demonstrate the suitability of ZELDA, hitherto proposed for non-common path aberrations, for the precise measurement of piston, tip, and tilt errors in segmented apertures within the diffraction-limited domain. Measurements of the segment alignment errors do not require changing the typical ZELDA set-up, but the signal estimators. ZELDA arises as a multitask sensor because it can simultaneously measure cophasing aberrations, segment figuring errors, and continuous wavefront aberrations due to optics misalignment \citep{NDIAYE16}.

Our closed-loop simulations show that within its capture range the sensor drives the segment piston, tip, and tilt to sub-nanometric residuals in a few iterations. However, its asymmetric capture range limits the convergence dynamic.
Nonetheless, it can be used as a stand-alone sensor for wavefront errors up to 120~nm RMS with a $95\%$ chance of phasing the system.
Improvements on the capture range are foreseen using either multiwavelength or coherence-based methods (see \citet[][]{MARTINEZ16} for a review of these techniques), or by implementing innovative FPM designs \citep[][]{JACKSON16}.

We study the phasing process from a statistical point of view. To our knowledge, the probabilistic behavior of the phasing process has not previously been explored in a systematic way. The proposed probabilistic models are independent of the phasing sensor architecture and only rely on the boundaries of the sensor capture range. 
These models provide the opportunity to predict the success of the phasing process as a function of the initial aberration values, and thus help in the risk analysis assessment for the telescope and the observing programs.
The statistical behavior of the convergence for a system subject to misalignment has also been investigated. It shows that the displacement of the FPM and the pupil impact the system in different manners but none of them is revealed as a drawback for the sensor. 

Additionally, we note that the simulated pupils used in our simulations are simplified because telescope apertures exhibit amplitude discontinuities (central obscuration, inter-segments space, and spiders) in addition to phase misalignment. 
In this context, 
the proposed estimators remain valid because the analysis is performed from a point by point measurement within the segment area. Potential partially obscured segments will present an inherent loss of signal that can be compensated by adjusting the spatial position and/or geometry of the analysis zone $\mathcal{H}$. 
{Only limited to piston, tip, and tilt in this study, a higher order of on-segment phase aberration can be considered to account for segment curvature or segment figure errors. 
Assuming that these higher order aberrations are taken into account in the calibration matrix, it does not preclude the system to converge. However, higher order modes will introduce additional cross-terms between them, possibly reducing the capture range to some extent. These effects are currently under study \citep[e.g., ][]{VIGAN2016}.}

Finally, the concept proposed in this paper will be tested in laboratory in the framework of the SPEED project \citep{MARTINEZ16b}, and compared with other cophasing systems \citep{JANIN2016, MARTINACHE13}.

\begin{acknowledgements}
P. Janin-Potiron is grateful to Airbus Defense and Space (Toulouse, France) and the R\'egion PACA (Provence Alpes C\^ote d\'{}Azur, France) for supporting his PhD fellowship. P. J-P would like to thank Y. Potiron for sharing his mathematical wisdom. M. N'Diaye acknowledges support from the European Research Council (ERC) through the KERNEL project grant \#683029 (PI: F. Martinache) and would like to thank F. Martinache for his support.
\end{acknowledgements}

\bibliography{biblio}

\begin{appendix}
\section{Analytical expression of the standard deviation over a segmented aperture}
\label{app:std}

In this appendix, we detail the calculation of the analytical expression for the standard deviation of wavefront errors in the context of hexagonal segmented apertures.

\begin{figure}[!ht]
\centering
\includegraphics[height=0.335\textwidth]{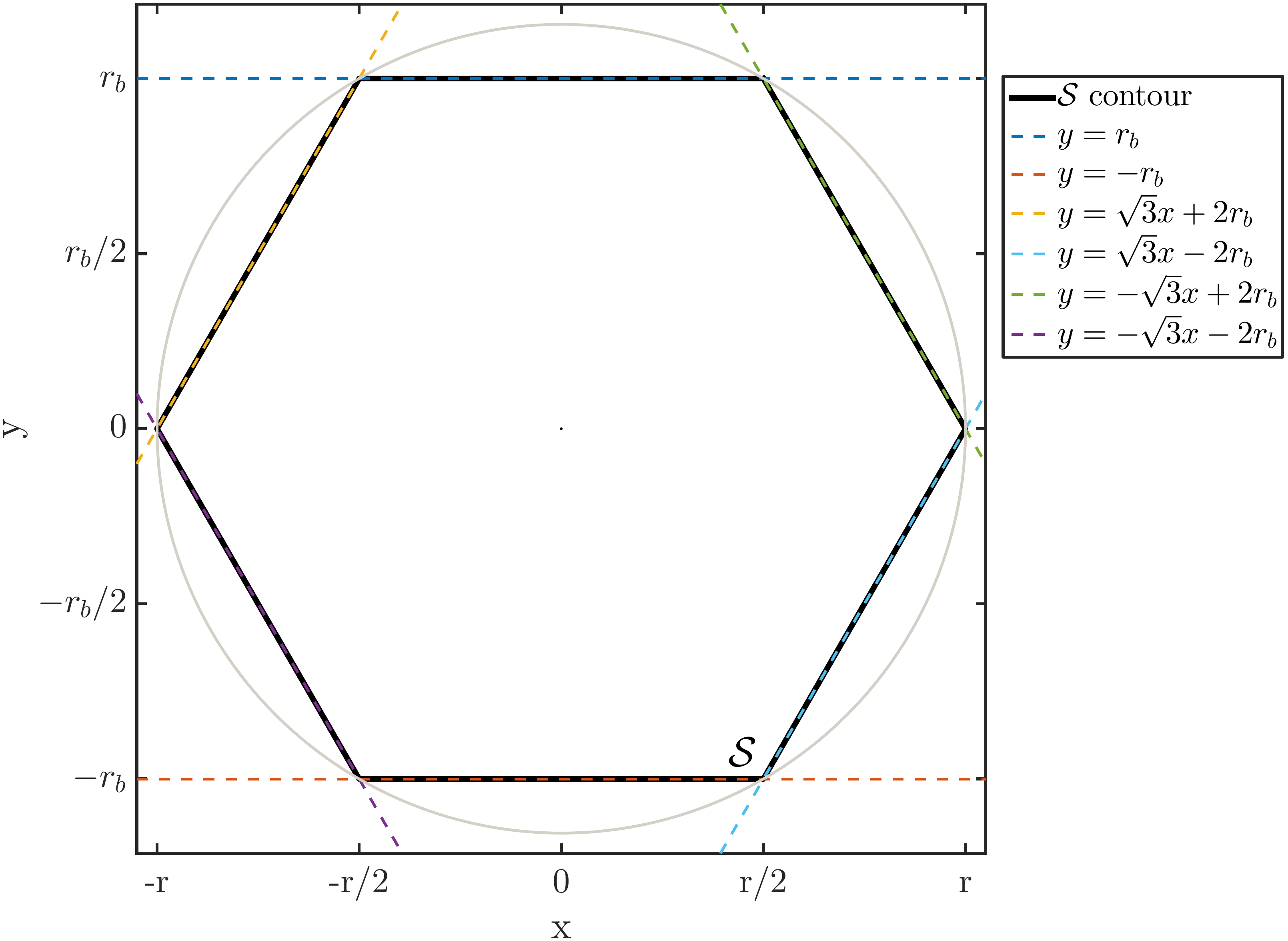}
\caption{Segment characteristics and boundary equations.}
\label{fig:segment}
\end{figure}

Let $h_n(x,y)$ be the elevation on the segment $n$ (with $\, n \in [1,N]$~) at the coordinates $(x,y)$,
\begin{equation}
h_n(x,y) = p_n + t_n .x + T_n .y \quad \forall (x,y) \in \mathcal{S},
\end{equation}
where $\mathcal{S}$ is the surface defined by a single segment as shown in Fig.~\ref{fig:segment}.
The expression of $\mathcal{S}$ is
\begin{equation}
\mathcal{S}=\left\{(x,y)\in \mathbb{R}^{\,2} \mid  x\in[-r,r], y\in[\zeta_1(x),\zeta_2(x)]\right\},
\end{equation} 
with
\begin{equation}
\zeta_1(x) = \left\{ \begin{array}{rcl}
-\sqrt{3}(x+r) & \mbox{for} & -r < x < -\frac{r}{2}\\
-r_b & \mbox{for} & -\frac{r}{2} \leq x \leq \frac{r}{2}\\
\sqrt{3}(x-r) & \mbox{for} & \frac{r}{2} < x < r\\
\end{array}\right.
\end{equation} 
and
\begin{equation}
\zeta_2(x) = \left\{ \begin{array}{rcl}
\sqrt{3}(x+r) & \mbox{for} & -r < x < -\frac{r}{2}\\
r_b & \mbox{for} & -\frac{r}{2} \leq x \leq \frac{r}{2}\\
-\sqrt{3}(x-r) & \mbox{for} & \frac{r}{2} < x < r.\\
\end{array}\right.
\end{equation}
\\
Hereafter, the notation $\overline{\mbox{...\raisebox{0.25cm}{}}}$ is used to represent the averaged value over one single segment, while $\mathbb{E}[...]$ is used to represent the averaged over the pupil or the average of the piston, tip, and tilt vectors. The mean elevation on the $n^\text{th}$ segment is
\begin{equation}
\begin{aligned}
\label{ref:meanhn}
\overline{h_n(x,y)}&= \frac{1}{A} \iint\limits_{\mathcal{S}} h_n(x,y) \mathrm{d} x \mathrm{d} y,
\\
&= \frac{1}{A} \left( p_n \iint\limits_{\mathcal{S}}  \mathrm{d} x \mathrm{d} y + t_n \iint\limits_{\mathcal{S}} x \mathrm{d} x \mathrm{d} y + T_n \iint\limits_{\mathcal{S}} y \mathrm{d} x \mathrm{d} y \right),
\end{aligned}
\end{equation}
where $A=3\sqrt{3}r^{\,2}/2$ represents the segment area. It leads to
\begin{equation}
\begin{aligned}
\overline{h_n(x,y)}&= p_n.
\end{aligned}
\end{equation}
The mean squared elevation on the $n^\text{th}$ segment is given by
\begin{equation}
\begin{aligned}
\overline{h_n^{\,2}(x,y)} &= \frac{1}{A} \iint\limits_{\mathcal{S}} h_n^{\,2}(x,y) \mathrm{d} x \mathrm{d} y ,
\\
&= \frac{1}{A} \iint\limits_{\mathcal{S}} \left(p_n^{\,2}+t_n^{\,2} x^{\,2} + T_n^{\,2} y^{\,2} + 2 p_n t_n x + \right.
\\
&\qquad\qquad \left. 2 p_n T_n y + 2 t_n T_n xy \right) \mathrm{d} x \mathrm{d} y.
\end{aligned}
\end{equation}
Developing {the} previous equation, we obtain
\begin{equation}
\begin{aligned}
\overline{h_n^{\,2}(x,y)} &= p_n^{\,2} + \frac{5}{24} r^{\,2} t_n^{\,2} + \frac{5}{24} r^{\,2} T_n^{\,2}.
\end{aligned}
\end{equation}
Let also $H$ be the union of all the segments composing the pupil:
\begin{equation}
H = \bigcup\limits_{n=1}^{N} h_n.
\end{equation}
The mean value of $H$ is
\begin{equation}
\mathbb{E}\left[H\right]= \frac{1}{N} \displaystyle\sum\limits_{n=1}^{N} \overline{h_n} = \mathbb{E}\left[\mathbf{p}\right].
\end{equation}
The mean value of $H^{\,2}$ is
\begin{equation}
\begin{aligned}
\mathbb{E}\left[H^{\,2}\right] &= \frac{1}{N} \sum\limits_{n=1}^{N} \overline{h_n^{\,2}}
\\
&= \mathbb{E}\left[\mathbf{p}^{\,2}\right] + \frac{5}{24} r^{\,2} \left(\mathbb{E}\left[\mathbf{t}^{\,2}\right] + \mathbb{E}\left[\mathbf{T}^{\,2}\right]\right).
\end{aligned}
\end{equation}
The variance over the full pupil is then expressed as
\begin{equation}
\begin{aligned}
\sigma_{pupil}^{\,2} &=\mathbb{E}\left[\left(H - \mathbb{E}\left[H\right] \right)^{\,2}  \right]= 
\mathbb{E}\left[H^{\,2}\right] - \mathbb{E}\left[H\right]^{\,2},
\\
&= \mathbb{E}\left[\mathbf{p}^{\,2}\right] - \mathbb{E}\left[\mathbf{p}\right]^{\,2} + \frac{5}{24} r^2\left(\mathbb{E}\left[\mathbf{t}^{\,2}\right] + \mathbb{E}\left[\mathbf{T}^{\,2}\right] \right).
\end{aligned}
\end{equation}
Finally, using the standard deviation of the piston, tip, and tilt samples, respectively $s_p$, $s_t,$ and $s_T$, we obtain
\begin{equation}
\begin{aligned}
\sigma_{pupil} = \sqrt{s_p^{\,2} + \frac{5}{24} r^2\left(s_t^{\,2} + s_T^{\,2} + \mathbb{E}\left[\mathbf{t}\right]^{\,2} + \mathbb{E}\left[\mathbf{T}\right]^{\,2} \right)}.
\end{aligned}
\end{equation}

\section{Impact of the ZELDA asymmetric capture range}
\label{app:comparison}

\begin{figure*}[!ht]
\centering
\includegraphics[height=0.38\textwidth]{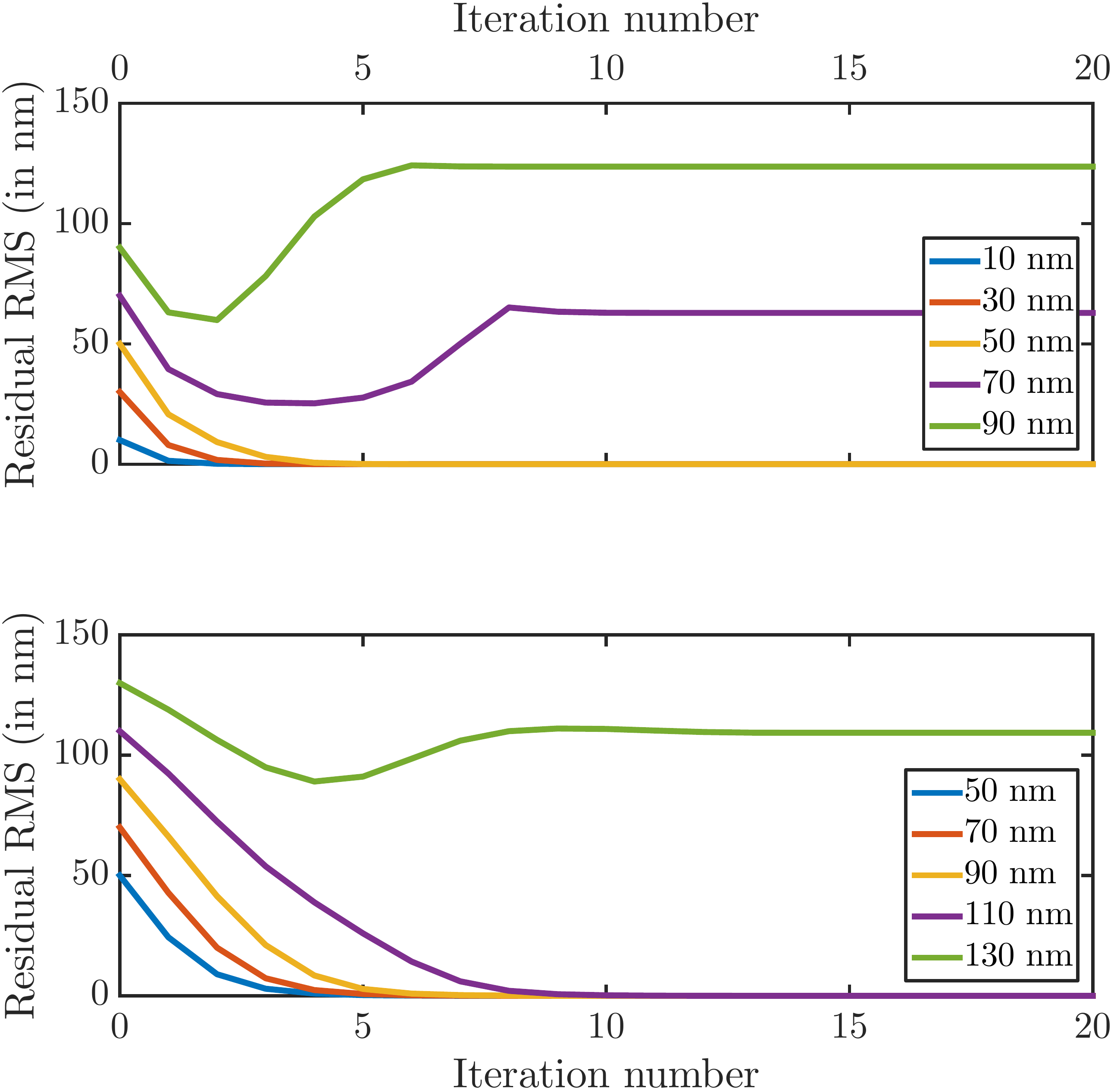}
\hspace{0.05\textwidth}
\includegraphics[height=0.38\textwidth]{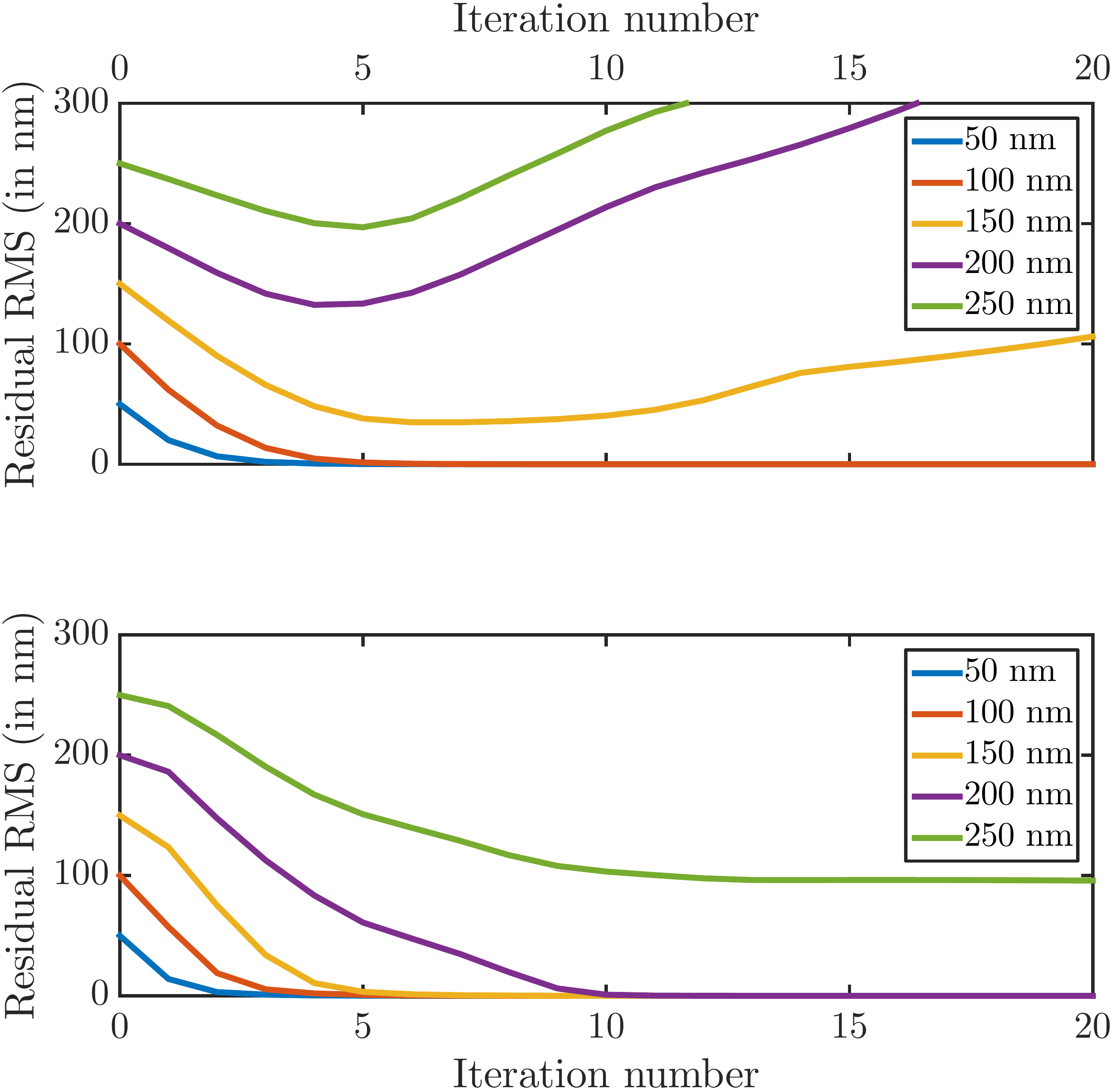}
\caption{Iterative evolution of the residual RMS wavefront error for piston only (left) and for combined piston and tip-tilt (right) with ZELDA (top) and the SCC-PS (bottom). Different initial conditions are considered.}
\label{fig:SCCZeldaPiston}
\end{figure*}

In this appendix, we propose to assess the impact of the asymmetry in the ZELDA capture range on the closed-loop convergence. The SCC-PS \citep{JANIN2016} has a symmetric piston capture range and is used as a reference for comparison purpose.
For both sensors, we use the same configuration and the same initial conditions as stated in Sec.~\ref{sec:closed-loop}.

In this piston-only configuration, the limit of convergence is reached for different initial wavefront errors: below $70$~nm RMS with ZELDA (Fig.~\ref{fig:SCCZeldaPiston}, top left), $\sim110$~nm RMS with a symmetric capture range sensor (Fig.~\ref{fig:SCCZeldaPiston}, bottom left).  
This reduction with ZELDA is a direct effect of the piston capture range shifting as described in Sec. \ref{subsec:analytical}.
For a larger amount of aberrations, the system converges toward a stable state with one (or more) segment(s) phased at a multiple of the wavelength.

In the piston plus tip-tilt configuration, the limit of convergence is also reached for different initial wavefront errors: below $\sim140$~nm RMS for ZELDA (Fig. \ref{fig:SCCZeldaPiston}, top right), $\sim200$ nm RMS with a symmetric capture range sensor (Fig. \ref{fig:SCCZeldaPiston}, bottom right).
For the symmetric capture range sensor (Fig. \ref{fig:SCCZeldaPiston}, top right, green curve), when the initial wavefront errors are too large, the system converges toward a stable non-phased configuration, while it diverges for ZELDA. This effect is intrinsic to the SCC-PS used here as a reference cophasing sensor, and will therefore not be discussed further.


\end{appendix}

\end{document}